\documentclass[aps,pra,twocolumn,floatfix,superscriptaddress,longbibliography,notitlepage]{revtex4-2}

\usepackage[utf8]{inputenc}
\usepackage[T1]{fontenc}
\usepackage{lmodern}
\usepackage{amsmath,amssymb,amsfonts,bm,mathrsfs}
\usepackage{graphicx}
\usepackage{microtype}
\usepackage{xcolor}
\usepackage[colorlinks=true,linkcolor=blue,citecolor=blue,urlcolor=blue]{hyperref}
\usepackage{orcidlink}

\newcommand{\Tr}{\operatorname{Tr}}

\newcommand{\ket}[1]{\left|#1\right\rangle}
\newcommand{\bra}[1]{\left\langle#1\right|}
\newcommand{\mel}[3]{\left\langle #1 \middle| #2 \middle| #3 \right\rangle}

\begin{document}

\title{Parity-selected charging and noise-correlated work retention in a coupled Si/SiGe 
double quantum dot battery: theory and experimentally parameterized modeling}
\author{Khalil Loukhssami~\orcidlink{0009-0008-7934-3001}}
\email{khalil.loukhssami@um5r.ac.ma}
\affiliation{ESMaR, Department of Physics, Faculty of Sciences, Mohammed V University in Rabat, Rabat, Morocco}

\author{Khadija El Hawary~\orcidlink{0000-0003-1204-3297}}
\email{khadija_elhawary@um5.ac.ma}
\affiliation{ESMaR, Department of Physics, Faculty of Sciences, Mohammed V University in Rabat, Rabat, Morocco}

\author{Sanaa Abaach~\orcidlink{0000-0003-3999-9045}}
\affiliation{ESMaR, Department of Physics, Faculty of Sciences, Mohammed V University in Rabat, Rabat, Morocco}

\author{Morad El Baz~\orcidlink{0000-0001-5201-2212}}
\affiliation{ESMaR, Department of Physics, Faculty of Sciences, Mohammed V University in Rabat, Rabat, Morocco}

\date{October 2026}

\begin{abstract}
We investigate the interplay between resonant charging, charge noise, and retention of extractable work in a quantum battery composed of two capacitively coupled silicon double quantum dots. By exploiting the symmetry of the interacting Hamiltonian, we identify a microwave transition selected by parity that enables efficient population transfer between the lowest and highest energy eigenstates. Using stochastic simulations of $1/f$ detuning noise with finite spectral bandwidth, we demonstrate that the stored ergotropy is predominantly governed by non-passive energy level populations, while quantum coherence accompanies the charging dynamics. Beyond charging, we establish how parity constrains phonon-induced relaxation and how spatial correlations of electrical fluctuations modify the competing leakage channels. The two electrical leakage channels exhibit opposite responses to spatial noise correlations, leading to enhanced retention of population ergotropy after the charging pulse under anticorrelated detuning fluctuations within the considered model. Our results reveal a direct connection between symmetry-controlled transitions, environmental noise correlations, and useful work preservation, providing a basis for assessing the stability of semiconductor quantum batteries.
\end{abstract}

\maketitle
\section{Introduction}
\label{sec:introduction}
Quantum batteries (QBs) provide a microscopic framework for studying energy storage, charging dynamics, and extractable work in finite-dimensional quantum systems. In contrast to classical batteries, where energy storage is governed by macroscopic electrochemical processes, QBs can store energy through coherent evolution, population redistribution, and correlations between elementary quantum units. Their useful performance is most naturally quantified by the ergotropy, namely the maximum work that can be extracted from a quantum state by unitary operations with respect to a fixed battery Hamiltonian. This distinction is essential because an increase in the mean energy does not necessarily imply an equivalent increase in extractable work~\cite{Allahverdyan2004,Alicki2013,Hovhannisyan2013,Campaioli2024}.
A central question in QB physics is how quantum resources contribute to useful work storage~\cite{Francica2020,Caravelli2021,Shi2022,Niu2024,Tirone2025NoisyQB}. Coherence and entanglement can modify the charging pathway, enhance power, or affect work extraction~\cite{Razzoli2025QST}, but they do not by themselves determine the full ergotropy. Recent theoretical and experimental studies have shown that ergotropy can be separated into coherent and incoherent (population) contributions. In this view, useful work may originate either from energy-basis coherences or from non-passive population distributions~\cite{Li2025CoherentErgotropy,Wang2026Ergotropy,loukhssami2026quantum}. This distinction is particularly important in driven multilevel batteries, where a coherent drive may first create superpositions and then redistribute populations in a way that violates passive ordering.

Semiconductor quantum dots (QDs) offer a promising solid-state platform for exploring such mechanisms because they combine strong confinement, fast electrical control, and compatibility with scalable semiconductor architectures~\cite{Zwanenburg2013,Burkard2023SemiQubits}. In particular, silicon/silicon-germanium (Si/SiGe) QDs are attractive because silicon-based architectures provide strong prospects for scalability and reduced nuclear spin noise compared with gallium arsenide (GaAs) platforms, while charge noise remains a central limitation for electrically controlled charge dynamics~\cite{Zwanenburg2013,Burkard2023SemiQubits,Mills2019Shuttling,petersson2010quantum}. In a double quantum dot (DQD), the localized charge states form an effective charge qubit, while the detuning and tunnel coupling provide two direct control axes. The detuning sets the relative energy of the localized charge configurations, whereas the tunnel coupling hybridizes them and opens an anticrossing. This makes DQDs natural platforms for coherent charge oscillations, microwave spectroscopy, Landau--Zener physics, and electrically driven charge qubit dynamics~\cite{Oosterkamp1998,Hayashi2003,petersson2010quantum,Kim2015,Shevchenko2010,Mi2018LZ}.
Beyond isolated DQDs, capacitively coupled DQDs provide an additional ingredient that is highly relevant for QB applications: an electrically tunable interaction between charge qubits. Experiments have demonstrated microwave-driven transitions, capacitive interactions, and correlated charge-qubit dynamics~\cite{Petersson2009,Ward2016,MacQuarrie2020}. Related theoretical studies have investigated quantum correlations in coupled semiconductor quantum-dot systems~\cite{mansour2020quantum,abaach2022long}. In the localized charge basis, the capacitive interaction acts as a conditional electrostatic shift and is naturally represented by an effective Ising-type term proportional to \(\sigma_z^{(1)}\sigma_z^{(2)}\). Such an interaction provides a minimal and experimentally motivated route to correlated charging dynamics in a quantum battery composed of two capacitively coupled DQDs.
The electric dipole moment that enables charge control also makes DQDs sensitive to electrostatic fluctuations from charge traps, interface disorder and other fluctuators, often characterized by a low-frequency $1/f^\alpha$ spectrum~\cite{Dutta1981,Weissman1988,Kogan1996,Schriefl2006TLF,Paladino2014,Beaudoin2015ChargeNoise,Szankowski2017NoiseSpectroscopy,Mickelsen2023}. These fluctuations primarily shift the detuning and can degrade charge transfer and electrically driven qubit operations~\cite{Dial2013,Freeman2016ChargeNoise,Petit2018HotSilicon,Connors2019,Connors2022,Krzywda2020}. Detuning and tunnel-coupling fluctuations need not have comparable strengths~\cite{Huang2018TunnelingNoise}; we therefore use tunnel-coupling modulation for charging and treat detuning fluctuations as the principal uncontrolled electrical perturbation.

The effect of $1/f$ detuning noise depends on its bandwidth relative to the microwave-pulse duration. Slowly varying components produce realization-dependent detuning and resonance shifts, whereas faster components perturb the driven evolution. A single phenomenological Markovian decay constant generally cannot reproduce these distinct contributions. The influence of environmental memory on ergotropy has also been examined in open-system battery models~\cite{Morrone2023QST}. Stochastic Hamiltonian methods retain the temporal correlations of the colored noise and permit the charging observables to be evaluated after ensemble averaging~\cite{Krzywda2020,Yang2019PRA}.
Fast charging through transverse-field control has been investigated in interacting two-qubit batteries~\cite{Evangelakos2025QST}. Microwave driving provides a possible route to robust operation because faster control can reduce the exposure time to slow charge noise and move the relevant dynamics toward higher frequencies where (1/f) noise is weaker. However, strong driving may also introduce additional effects, including Bloch--Siegert shifts, fast oscillations, leakage, and noise-induced fluctuations of the Rabi frequency. Studies of strongly driven semiconductor quantum-dot hybrid qubits have examined noise sensitivity and high-fidelity control~\cite{Yang2019PRA}. More generally, pulse-shaping methods can reduce leakage and improve frequency selectivity in driven quantum systems~\cite{Motzoi2009,Motzoi2013}. These results motivate the search for charging protocols that combine fast coherent transfer with reduced sensitivity to detuning noise.
Although charge noise has been extensively investigated in semiconductor qubits, its implications for the formation and subsequent retention of extractable work in coupled DQD batteries remain of interest. Experimental work on superconducting quantum batteries has also documented self-discharge after charging~\cite{Hu2022QST}. The separation of coherent and population ergotropy under noise has already been considered~\cite{Francica2020,Arjmandi2026NoiseWork}; our emphasis is therefore on the parity-selected charging pathway and on the environmental channels controlling work retention. In particular, lattice phonons and electrical fluctuations act through distinct couplings~\cite{Krzywda2020,Connors2022}, while spatially correlated detuning noise can alter the relative strengths of the transitions connecting the charged state to intermediate levels.

In this work, we develop a noise-aware model of a quantum battery based on two capacitively coupled Si/SiGe DQDs with experimentally motivated coherent parameters. We first study resonant microwave charging by tunnel-coupling modulation at the symmetric zero-detuning operating point. The charge fluctuations are represented by time-dependent stochastic traces with a band-limited $1/f$ spectrum. For each realization, the battery evolves unitarily, whereas the ensemble-averaged state describes the observed loss of coherent contrast and resonant population transfer. We calculate the total ergotropy, its population and coherent contributions, the energy-basis populations, the passive-ordering violation indicator, and Jensen--Shannon coherence. An exact parity analysis selects the dominant $E_0\leftrightarrow E_3$ transition. The results show that the resonant field generates coherence during population transfer, while the dominant part of the extractable work is stored in non-passive population distributions. Detuning noise reduces this work primarily through the weakening of the population-transfer pathway as well as the accompanying coherence.

We then examine population-ergotropy retention after a common ideal charging pulse. To distinguish storage from charging, the microwave drive is switched off before the dynamics with acoustic relaxation is initiated. A microscopic deformation-potential model of silicon phonons is combined with spatially correlated detuning fluctuations to evaluate the channels through which the charged state evolves. Parity constrains the longitudinal phonon matrix elements, while electrical cross correlations change the relative strengths of intermediate-level leakage channels. We relate these channel-dependent effects to the retained population ergotropy computed from the noise-averaged density matrix. For the specified orbital geometry and noise spectrum, anticorrelated fluctuations preserve more population ergotropy over the investigated interval than independent or positively correlated fluctuations. A complementary finite-temperature analysis in the Appendix provides an analytical benchmark for passive ordering and the available extractable work. The post-pulse calculations concern the retention of a common ideally charged state; phonon dissipation during the microwave pulse is outside the present model.

\section{Model and methods}
\label{sec:model_methods}

\subsection{Static battery Hamiltonian}
\label{subsec:static_hamiltonian}

We consider a quantum battery composed of two capacitively
coupled DQDs, each operating in the
charge-qubit regime. The low-energy subspace of each DQD
is spanned by the localized states $\ket{L}$ and $\ket{R}$,
corresponding to an excess electron occupying the left
or right dot. The coupled system is described in the
product basis
$\{\ket{LL},\ket{LR},\ket{RL},\ket{RR}\}$.
These charge configurations provide the standard basis
for coherent charge oscillations and electrically driven
dynamics in semiconductor DQDs
~\cite{petersson2010quantum,Kim2015}.

The static battery Hamiltonian is
\begin{equation}
H_B =
\sum_{i=1}^{2}
\left[
\frac{\epsilon_{0,i}}{2}\sigma_z^{(i)}
+
\frac{\Delta_{0,i}}{2}\sigma_x^{(i)}
\right]
+
\frac{J}{4}\sigma_z^{(1)}\sigma_z^{(2)} .
\label{eq:HB_theory}
\end{equation}
Here, $\epsilon_{0,i}$ denotes the static detuning of
the $i$th DQD, while $\Delta_{0,i}=2t_c^{(i)}$ is its
tunnel splitting, with $t_c^{(i)}$ the interdot tunnel
coupling. The longitudinal term controls the charge
polarization, whereas the transverse term hybridizes
the localized states and opens an avoided crossing
near zero detuning. The Ising interaction $J$ produces
a conditional electrostatic shift determined by the
joint charge configuration of the two DQDs. As a
result, the interacting spectrum contains correlated
energy levels and transition frequencies that cannot
be obtained by simply adding the spectra of two
independent charge qubits
~\cite{MacQuarrie2020}.
The eigenstates of $H_B$ are denoted by
$\ket{\phi_n}$, with the corresponding eigenenergies
ordered as $E_0\leq E_1\leq E_2\leq E_3$.
This fixed energy eigenbasis is used throughout the
analysis to define populations, coherences, passive
states, and extractable work.

Unless otherwise stated, we adopt the experimentally
motivated Si/SiGe parameters
$\Delta_{0,1}/h=8.4\,\mathrm{GHz}$,
$\Delta_{0,2}/h=6.6\,\mathrm{GHz}$, and
$J/h=15.3\,\mathrm{GHz}$, based on the capacitively
coupled charge-qubit device of
Ref.~\cite{MacQuarrie2020}.
In the convention of that reference, the capacitive
interaction is written as
$H_{\mathrm{int}}=
g(\mathbb I-\sigma_z^{(1)})
(\mathbb I-\sigma_z^{(2)})/4$.
Up to an irrelevant scalar energy shift, this
corresponds to $J=g$, with the induced local shifts
absorbed into
$\epsilon_{0,i}=\epsilon_{i,\mathrm{gate}}-g/2$.
The adopted experimental parameters set the coherent
energy scales but do not independently determine the
orbital geometry or phonon relaxation rates.

We choose the symmetric operating point
$\epsilon_{0,1}=\epsilon_{0,2}=0$.
For an isolated charge qubit, zero detuning corresponds
to the charge degeneracy point, where the energy
splitting is first-order insensitive to static
detuning fluctuations~\cite{Yang2019PRA}.
In the coupled system, this operating point also
supports an exact parity symmetry of $H_B$, which
constrains the allowed transitions under microwave
driving. The comparable tunnel and capacitive energy
scales produce a well-resolved interacting spectrum
with selectively addressable transitions.

The symmetric point reduces sensitivity to static
detuning perturbations but does not eliminate the
effects of time-dependent charge noise during resonant
driving. In particular, stochastic detuning fluctuations
can modify the population-transfer dynamics and
couple different parity sectors. This operating point
should therefore not be identified with an optimized
alternating-current (AC) sweet spot, which would require additional
conditions on the driven control parameters.
The corresponding symmetry selection rules and noise
effects are examined in the following subsections.

\subsection{Microwave charging through tunnel coupling modulation}
\label{subsec:Adelta_drive}

The charging protocol is implemented by resonantly modulating the tunnel coupling channel. This choice is motivated by two complementary considerations. First, the tunnel coupling controls the hybridization between $\ket{L}$ and $\ket{R}$ and therefore provides a transverse control axis for inducing transitions between eigenstates of $H_B$. Second, the dominant uncontrolled electrical noise in gate-defined DQDs mainly enters through the detuning coordinate, because detuning fluctuations shift the relative energy of the localized charge states \cite{Paladino2014,Connors2019,Connors2022}. Driving through the tunnel coupling channel therefore injects energy into the battery without intentionally modulating the noisiest control parameter.
During the charging stage, the deterministic Hamiltonian is
\begin{equation}
    H_0(t)=H_B+H_{\mathrm{ch}}(t),
    \label{eq:H0_theory}
\end{equation}
with
\begin{equation}
    H_{\mathrm{ch}}(t)
    =
    \sum_{i=1}^{2}
    \frac{A_{\Delta,i}}{2}
    \cos\!\left(\omega_d t+\varphi_i\right)
    \sigma_x^{(i)} .
    \label{eq:Hcharging_Adelta}
\end{equation}
Here, $A_{\Delta,i}$ is the modulation amplitude of the tunnel splitting of DQD $i$, $\omega_d$ is the angular drive frequency, and $\varphi_i$ is the drive phase. Unless otherwise stated, we consider symmetric, in-phase driving, $A_{\Delta,1}=A_{\Delta,2}\equiv A_\Delta$ and $\varphi_1=\varphi_2=0$, so that
\begin{equation}
    H_{\mathrm{ch}}(t)
    =
    \frac{A_\Delta}{2}
    \cos(\omega_d t)
    \left(
        \sigma_x^{(1)}+\sigma_x^{(2)}
    \right).
    \label{eq:Hcharging_symmetric}
\end{equation}
This term is transverse in the localized charge basis. It does not shift the localized charge states directly; instead, it modulates the interdot hybridization and couples the eigenstates of the interacting battery Hamiltonian. Such transverse microwave control is closely related to the strong driving protocols developed for charge and hybrid QD qubits \cite{thorgrimsson2017extending}.

\paragraph{Parity-selected charging doublet.}
At the symmetric operating point $\epsilon_{0,1}=\epsilon_{0,2}=0$,
$\Pi=\sigma_x^{(1)}\sigma_x^{(2)}$ is an exact symmetry of the static
Hamiltonian and symmetric driving operator $D=\sigma_x^{(1)}+\sigma_x^{(2)}$:
\begin{equation}
\Pi^2=\mathbb I,\quad[H_B,\Pi]=[D,\Pi]=0,\quad
\{\Pi,\sigma_z^{(i)}\}=0.
\label{eq:parity_exact}
\end{equation}
In the even ($+$) and odd ($-$) parity subspaces, the two positive
energy scales are
\begin{equation}
\Lambda_\pm=\sqrt{\frac{(\Delta_{0,1}\pm\Delta_{0,2})^2}{4}
+\frac{J^2}{16}}.
\label{eq:parity_energies}
\end{equation}
For positive tunnel splittings the energy-ordered states obey
$E_0=-\Lambda_+$, $E_1=-\Lambda_-$, $E_2=\Lambda_-$,
$E_3=\Lambda_+$. States $\phi_0,\phi_3$ have parity $+1$, whereas
$\phi_1,\phi_2$ have parity $-1$. Since the microwave control preserves
parity, $\langle\phi_m|D|\phi_n\rangle=0$ for opposite parity.
For the selected branch,
\begin{equation}
|M_{30}^{(\Delta)}|=\frac{|J|}{2\Lambda_+},\qquad
f_{03}=\frac{2\Lambda_+}{h}.
\label{eq:03_exact}
\end{equation}
Consequently, the $E_0\leftrightarrow E_3$ charging pathway is symmetry selected and remains decoupled from the opposite-parity sector under ideal tunnel modulation. This selection rule does not imply immunity to environmental perturbations, since detuning noise anticommutes with $\Pi$ and can open transitions towards $E_1$ and $E_2$. The strength of these leakage channels is assessed in the post-pulse analysis.

The drive frequency is chosen from the static many-body spectrum. A transition $\ket{\phi_n}\leftrightarrow\ket{\phi_m}$ is efficiently addressed when 

\begin{equation}
    \hbar\omega_d\simeq E_m-E_n
    \label{eq:resonance_condition}
\end{equation}
and when the corresponding tunnel drive matrix element
\begin{equation}
    M_{mn}^{(\Delta)}
    =
    \bra{\phi_m}
    \left(
        \sigma_x^{(1)}+\sigma_x^{(2)}
    \right)
    \ket{\phi_n}
    \label{eq:tunnel_matrix_element}
\end{equation}
is sufficiently large. Since the initial Gibbs state at $20\,\mathrm{mK}$ overwhelmingly occupies $\ket{\phi_0}$, the relevant dominant charging channels are the transitions $\ket{\phi_0}\rightarrow\ket{\phi_m}$ that combine a large energy gap with a large drive matrix element. In the main protocol, the microwave is tuned close to the dominant transition
\begin{equation}
    f_d=\frac{E_3-E_0}{h},
    \qquad
    \omega_d=2\pi f_d .
    \label{eq:drive_frequency}
\end{equation}
This resonant protocol transfers population from the low-energy sector to the high-energy sector of the battery while generating energy-basis coherences. The charging process is therefore neither a purely classical population pump nor a purely coherence storage process. Rather, the microwave field first creates coherent superpositions, and the subsequent driven dynamics can convert part of this coherence into a non-passive population distribution. This mechanism is important for QB operation because ergotropy can arise from both off-diagonal coherences and violations of passive population ordering in the energy basis \cite{Francica2020,Niu2024,Wang2026Ergotropy}.

In the absence of charge noise, the density matrix evolves according to
\begin{equation}
    \frac{d\rho_0(t)}{dt}
    =
    -\frac{i}{\hbar}
    \left[
        H_0(t),\rho_0(t)
    \right],
    \label{eq:rho_noise_free}
\end{equation}

\paragraph{Thermal initial preparation.}
The initial condition throughout Figs.~\ref{fig:spectrum_dynamics}--\ref{fig:ergotropy_maps}, and for the common noiseless
charged state used in Fig.~\ref{fig:phonon_experimental_audit}, is the Gibbs state of the same static battery
Hamiltonian,
\begin{align}
\rho_\beta&=\frac{e^{-\beta H_B}}{Z},\qquad
 Z=\Tr(e^{-\beta H_B}),\label{eq:initial_gibbs_new}\\
\beta&=\frac{1}{k_B T_{\mathrm{init}}},\qquad
 T_{\mathrm{init}}=20\,\mathrm{mK}.
\end{align}
The energy-basis populations are
\begin{equation}
 p_n^\beta=\frac{e^{-(E_n-E_0)/(k_B T_{\mathrm{init}})}}
 {\sum_m e^{-(E_m-E_0)/(k_B T_{\mathrm{init}})}}.
 \label{eq:gibbs_populations_model}
\end{equation}
At the symmetric point, \((E_1-E_0)/k_B\simeq215\,\mathrm{mK}\), so
\(p_0^\beta\simeq0.999979\) and \(p_1^\beta\simeq2.1\times10^{-5}\).
This low-temperature preparation is nearly ground-state dominated but is
propagated as a mixed state. It is a theoretical benchmark rather than the
measured electronic temperature \(T_e\simeq228\,\mathrm{mK}\) reported in
Ref.~\cite{MacQuarrie2020}; the ideal Gibbs-temperature dependence is
examined separately in Appendix~\ref{app:temperature_20_228}.
The phonon bath in the post-pulse study is taken at
\(T_{\mathrm{ph}}=100\,\mathrm{mK}\), independently of
\(T_{\mathrm{init}}\). The noiseless density operator satisfies
\(\rho_0(0)=\rho_\beta\); its evolution defines the reference charging
protocol before stochastic fluctuations are introduced.

In the following, detuning fluctuations are introduced as a stochastic perturbation, and the resulting noise-averaged state is compared with this noiseless reference to quantify the degradation of population transfer, energy-basis coherence, and ergotropy.

\subsection{Detuning charge noise}
\label{subsec:detuning_noise}

Electrostatic fluctuations from charge traps, interface disorder and fluctuating two-level systems shift the energy bias between localized DQD charge states~\cite{Dutta1981,Weissman1988,Kogan1996,Dial2013,Connors2019}. Their detuning coupling motivates the stochastic longitudinal perturbation used below.

In the present battery protocol, this mechanism plays a particularly important role. The coherent charging field is applied through the tunnel coupling channel, whereas the dominant uncontrolled electrical noise is assumed to perturb the detuning coordinates. We therefore write the instantaneous detuning of the $i$th DQD as $\epsilon_i(t)=\epsilon_{0,i}+\delta\epsilon_i(t)$, where $\epsilon_{0,i}$ is the chosen static operating point and $\delta\epsilon_i(t)$ is a classical stochastic fluctuation. The corresponding noise Hamiltonian is \cite{Paladino2014,Mickelsen2023}
\begin{equation}
H_n(t)=\sum_{i=1}^{2}\frac{\delta\epsilon_i(t)}{2}\sigma_z^{(i)} .
\label{eq:noise_hamiltonian}
\end{equation}
Thus, detuning noise is longitudinal in the localized charge basis. However, its effect on the battery dynamics is not limited to a simple phase fluctuation. Since the static Hamiltonian contains tunnel coupling terms proportional to $\sigma_x^{(i)}$, and since the system is driven during the charging stage, $H_n(t)$ does not generally commute with $H_B+H_{\mathrm{ch}}(t)$. Consequently, detuning noise can suppress energy-basis coherences, shift the resonance condition, and modify the population transfer pathway between the eigenstates of the interacting battery Hamiltonian.

For a given noise realization $\alpha$, the total Hamiltonian is
\begin{equation}
H_{\alpha}(t)=H_B+H_{\mathrm{ch}}(t)+\sum_{i=1}^{2}\frac{\delta\epsilon_i^{(\alpha)}(t)}{2}\sigma_z^{(i)} .
\label{eq:total_stochastic_hamiltonian}
\end{equation}
The stochastic processes $\delta\epsilon_i(t)$ are taken to have zero mean and stationary correlations. Unless otherwise stated, the two local detuning noises are considered independent and statistically identical. This assumption provides the minimal model for local charge noise-induced degradation. The independent-noise case defines the baseline for Figs.~\ref{fig:spectrum_dynamics}--\ref{fig:ergotropy_maps}; correlated-noise scenarios are considered separately in Fig.~\ref{fig:phonon_experimental_audit}.
Following standard models of low-frequency charge noise in semiconductor nanostructures, we describe the detuning fluctuations by a band-limited $1/f$ spectrum \cite{Kogan1996,Paladino2014,Yang2019PRA}. Using the angular frequency convention, the noise power spectral density (PSD) inside the finite bandwidth is written as
\begin{equation} \tilde{S}_{\epsilon}(\omega)= \begin{cases} \dfrac{2\pi c_{\epsilon}^{2}}{|\omega|}, & \omega_l\leq |\omega|\leq \omega_h,\\[0.7em] 0, & \mathrm{otherwise}. \end{cases} \label{eq:one_over_f_noise} \end{equation}
Here, $c_{\epsilon}$ controls the spectral noise strength, while $\omega_l$ and $\omega_h$ are the low and high angular frequency cutoffs defining the finite noise bandwidth. Throughout the noise model we use angular frequencies, with $\omega=2\pi f$.
It is important to distinguish the spectral amplitude $c_{\epsilon}$ from the time-domain
root-mean-square (RMS) detuning fluctuation. With the two-sided angular-frequency convention used in Eq.~\eqref{eq:one_over_f_noise}, the RMS fluctuation integrated over the finite bandwidth is
\begin{equation}
\sigma_{\epsilon}^{2}
=
2c_{\epsilon}^{2}\ln(\omega_h/\omega_l).
\label{eq:rms_detuning_noise}
\end{equation}
Thus, $c_{\epsilon}$ is the parameter that fixes the amplitude of the band-limited $1/f$ spectrum used to generate the stochastic detuning traces, whereas $\sigma_{\epsilon}$ is a bandwidth-dependent RMS quantity obtained after integrating that spectrum. In the figures below, the noise strength is therefore labeled by $c_{\epsilon}$, which is the scanned simulation parameter, and not by $\sigma_{\epsilon}$. The latter is quoted only as an equivalent RMS scale for the chosen frequency window.
For $c_\epsilon=0.6\,\mu\mathrm{eV}$ and the 1 Hz--256 GHz window, Eq.~\eqref{eq:rms_detuning_noise} gives $\sigma_\epsilon\simeq 4.35\,\mu\mathrm{eV}$. This value counts the entire band, including the quasi-static part, and is not the amplitude of an additional independent disorder distribution.\par
We do not assign a universal value to $c_{\epsilon}$, since charge noise amplitudes depend strongly on the material stack, device geometry, gate configuration, and operating point. Instead, $c_{\epsilon}$ is treated as a tunable spectral parameter that allows us to quantify the robustness of the charging protocol over a controlled range of detuning noise strengths.

The low-frequency cutoff is much smaller than the inverse nanosecond pulse duration. Consequently, the part of the specified spectrum that cannot be resolved as an oscillation within a finite numerical record is represented by a quasi-static random Fourier-band component. Its variance is obtained by integrating that same spectrum over the unresolved interval, rather than appending an independent disorder distribution. The remaining modes are generated by Fourier synthesis: amplitudes follow Eq.~\eqref{eq:one_over_f_noise}, random phases generate independent traces, and the inverse transform gives $\delta\epsilon_i^{(\alpha)}(t)$. Each realization is propagated under $H_\alpha(t)$ before the density matrices are averaged. This construction preserves the distinction between static offsets and dynamical fluctuations without counting the same spectral weight twice. The Gaussian model approximates the combined effect of many weak fluctuators; a single strongly coupled fluctuator need not be Gaussian~\cite{Paladino2014,Krzywda2020}. Spectral normalization, finite-record resolution and time discretization are addressed in Appendix~\ref{app:phonon_repro_added}.

\subsection{Cross-correlated $1/f$ detuning noise}
\label{sec:crossnoise_new}
Nearby Si quantum dots can experience common electrical fluctuators;
measurable correlations have been found in neighboring silicon spin qubits
\cite{Yoneda2023}. To test their impact in the present charge-qubit
model, replace the independent local spectra by the matrix
\begin{equation}
\mathbf S_\epsilon(\omega)=\widetilde S_\epsilon(\omega)
\begin{pmatrix}1&\kappa\\\kappa&1\end{pmatrix},
\quad -1\le\kappa\le1.
\label{eq:noise_matrix_new}
\end{equation}
The eigenvalues $\widetilde S_\epsilon(1\pm\kappa)$ ensure a
nonnegative covariance spectrum. The case $\kappa=0$ recovers the
original results; $\kappa=\pm1$ represent perfectly correlated and
anticorrelated noise. A constant $\kappa$ over the full bandwidth
is a controlled modeling choice, not an experimental measurement.
For independent unit-normalized $1/f$ traces $u_{1,2}$, an explicit
construction is
\begin{align}
\delta\epsilon_1(t)&=c_\epsilon u_1(t),\\
\delta\epsilon_2(t)&=c_\epsilon
[\kappa u_1(t)+\sqrt{1-\kappa^2}\,u_2(t)],
\label{eq:noise_sampling_new}
\end{align}
with normalization chosen to reproduce the precise PSD convention of
Eq.~\eqref{eq:one_over_f_noise}. With
$\delta\epsilon_\pm=(\delta\epsilon_1\pm\delta\epsilon_2)/2$,
\begin{equation}
H_n=\frac{\delta\epsilon_+}{2}(\sigma_z^{(1)}+\sigma_z^{(2)})
+\frac{\delta\epsilon_-}{2}(\sigma_z^{(1)}-\sigma_z^{(2)}),
\label{eq:common_diff_new}
\end{equation}
and $S_\pm=\widetilde S_\epsilon(1\pm\kappa)/2$.
Distinct transition matrix elements for the two operators imply that
the dependence of leakage and population ergotropy on $\kappa$ must
be computed, rather than assumed. 
For a weak stationary classical detuning perturbation, define
\(Z_{n3}^{(i)}=\langle\phi_n|\sigma_z^{(i)}|\phi_3\rangle\).
The corresponding symmetric upward/downward transition-rate diagnostic is
\begin{equation}
\Gamma_{3\rightarrow n}^{1/f}(\kappa)=
\frac{\widetilde S_\epsilon(\omega_{3n})}{4\hbar^2}
\left[\sum_{i=1}^2|Z_{n3}^{(i)}|^2+
2\kappa\operatorname{Re}\bigl(Z_{n3}^{(1)}Z_{n3}^{(2)*}\bigr)\right].
\label{eq:electrical_leak_exact}
\end{equation}
Here \(\omega_{3n}=(E_3-E_n)/\hbar\).
These rates are used only to interpret the spectral leakage channels;
they are \emph{not} added to the explicitly sampled stochastic-Hamiltonian
dynamics, which already contains the same classical fluctuations.

\subsection{Silicon acoustic phonons: microscopic coupling and experimental constraints}
\label{sec:phonons_new}
The stochastic dynamics generated by the Hamiltonian
of Eq.~\eqref{eq:total_stochastic_hamiltonian} accounts for loss of
coherent contrast after averaging over classical charge-noise realizations,
but does not describe irreversible phonon emission and absorption.
For silicon, the leading long-wavelength acoustic interaction is the
\emph{deformation-potential} coupling, involving longitudinal and
transverse acoustic branches. The piezoelectric mechanism prominent in
GaAs does not provide the corresponding bulk-silicon channel
\cite{StavrouHu2005,KrzywdaCywinski2021Phonons}.
A phonon bath independent of the prescribed classical detuning noise is
introduced through
\begin{align}
H_{\mathrm{tot}}^{(\alpha)}(t)&=H_\alpha(t)+H_{\mathrm{ph}}+H_{\mathrm{ep}},\\
H_{\mathrm{ph}}&=\sum_{\mathbf q\lambda}\hbar\omega_{\mathbf q\lambda}
  b_{\mathbf q\lambda}^{\dagger}b_{\mathbf q\lambda},\\
H_{\mathrm{ep}}&=\sum_{\mathbf q\lambda}
 \left(b_{\mathbf q\lambda}+b_{-\mathbf q\lambda}^{\dagger}\right)
  M_{\mathbf q\lambda}\sum_{i=1}^2F_i(\mathbf q),
\label{eq:phonon_density_project}
\end{align}
where $\omega_{\mathbf q\lambda}=v_\lambda|\mathbf q|$,
and $M_{\mathbf q\lambda}$ includes the deformation potentials,
phonon polarization, mass density and mode normalization. This form
allows both DQDs to couple to common lattice modes.

For the illustrative calculation of Fig.~\ref{fig:phonon_experimental_audit},
we explicitly adopt an isotropic acoustic dispersion and a [001]-valley
single-axis deformation-potential vertex,
\begin{equation}
M_{\mathbf q\lambda}=i\sqrt{\frac{\hbar q}{2\rho_m Vv_\lambda}}
\bigl[\Xi_d(\hat{\bf q}\cdot\mathbf e_\lambda)
+\Xi_u(\hat{\bf q}\cdot\hat{\bf z})(\mathbf e_\lambda\cdot\hat{\bf z})\bigr].
\label{eq:material_dp_vertex_added}
\end{equation}

We use the representative bulk-silicon material parameters
$\rho_m=2330\,\mathrm{kg\,m^{-3}}$, $v_L=9330\,\mathrm{m\,s^{-1}}$,
$v_T=5420\,\mathrm{m\,s^{-1}}$, $\Xi_d=5.0\,\mathrm{eV}$, and
$\Xi_u=8.77\,\mathrm{eV}$. The deformation-potential values are employed
in the microscopic Si quantum-dot treatment of Gamble \emph{et al.}
\cite{Gamble2012SiPhonon}; the listed sound speeds and density are
effective isotropic acoustic inputs informed by the established
crystalline-silicon elastic properties~\cite{Hopcroft2010Elasticity},
rather than measurements on the Si/SiGe four-dot device. Experimental studies have also reported different
uniaxial deformation-potential estimates
\cite{Li1991XiExperimental}, highlighting their convention- and
sample-dependence. We use these parameters as a reproducible
isotropic-acoustic benchmark, without fitting phonon relaxation data;
crystal anisotropy, valley admixture, and actual confinement geometry
remain sources of device-specific uncertainty.

In each localized
orbital pair $\{|L_i\rangle,|R_i\rangle\}$, let
$F_{ab}^{(i)}(\mathbf q)=\langle a_i|e^{i\mathbf q\cdot\mathbf r}|b_i\rangle$.
The two-state projection is
\begin{equation}
F_i(\mathbf q)=f_{i0}\mathbb I+f_{ix}\sigma_x^{(i)}
+f_{iy}\sigma_y^{(i)}+f_{iz}\sigma_z^{(i)},
\label{eq:orbital_form_factor_pauli}
\end{equation}
with $f_{iz}=(F_{LL}^{(i)}-F_{RR}^{(i)})/2$,
$f_{ix}=(F_{LR}^{(i)}+F_{RL}^{(i)})/2$ and
$f_{iy}=(F_{RL}^{(i)}-F_{LR}^{(i)})/(2i)$.
The identity term cannot cause transitions within this truncated
charge subspace. The diagonal-density difference yields a longitudinal
$\sigma_z$ channel, while interorbital overlaps generate transverse
$\sigma_x$ and, in general, $\sigma_y$ channels. Their spectral weights
must be calculated from \emph{orthogonalized} localized orbitals:
introducing an independent transverse rate without its orbital form factor
would not constitute a microscopic prediction
\cite{KrzywdaCywinski2021Phonons}.

At zero static detuning the operator
$\Pi=\sigma_x^{(1)}\sigma_x^{(2)}$ commutes with $H_B$.
For a purely longitudinal projected coupling,
$\{\Pi,\sigma_z^{(i)}\}=0$ and
$\langle\phi_0|\sigma_z^{(i)}|\phi_3\rangle=0$.
The direct $E_3\to E_0$ \emph{single-phonon} transition is then
forbidden at leading order, although $E_3\to E_{1,2}$ and
$E_{1,2}\to E_0$ can remain allowed. By contrast,
$[\Pi,\sigma_x^{(i)}]=0$ and
$\langle\phi_0|\sigma_x^{(i)}|\phi_3\rangle$ need not vanish.
The magnitude of a transverse decay contribution is controlled by
$f_{ix}$, not by the system matrix element alone. Detuning away from the
symmetric point also lifts the longitudinal selection rule, so the
operating-point dependence must be retained.

For a static Hamiltonian, downward transition rates may be written as
\begin{align}
\Gamma_{m\to n}^{\mathrm{ph}}&=\frac{2\pi}{\hbar^2}
\sum_{\mathbf q\lambda}|M_{\mathbf q\lambda}|^2
\left|\left\langle\phi_n\left|
 \sum_iF_i(\mathbf q)\right|\phi_m\right\rangle\right|^2
\nonumber\\[-2pt]
&\quad\times[n_B(\omega_{mn},T)+1]
\delta(\omega_{mn}-\omega_{\mathbf q\lambda}),
\label{eq:phonon_rate_microscopic}
\end{align}
for $E_m>E_n$, $\omega_{mn}=(E_m-E_n)/\hbar$ and
$n_B=(e^{\hbar\omega/(k_BT)}-1)^{-1}$. Upward rates follow the
corresponding Bose occupation and thermal detailed balance.
These rates depend on dot separation, confinement lengths, crystallographic
axes, valley composition and the actual orbital wavefunctions. Neither
$\Delta_{0,i}/h$ nor $J/h$ uniquely fixes this geometry.
Mode interference and the cross-spectrum of the shared lattice bath must
be evaluated before replacing the bath by independent local jump channels.

The periodic tunnel drive requires additional care. With Floquet
quasienergies $\varepsilon_a$ and harmonics $k$, the dissipative
frequencies are
\begin{equation}
\omega_{ab}^{(k)}=(\varepsilon_b-\varepsilon_a)/\hbar+k\omega_d.
\label{eq:phonon_floquet_new}
\end{equation}
The corresponding phonon spectra must be sampled at these sidebands,
using the properly projected Floquet matrix elements
\cite{GrifoniHanggi1998}. A dissipator constructed only from the
unperturbed, noise-free Floquet modes is a \emph{reference approximation}
when slow detuning noise changes the instantaneous spectrum. In the present
work a static-basis dissipator is applied only after the charging pulse
has been switched off, within weak-coupling and secular approximations.
A driven Floquet--Markov treatment is beyond the present storage protocol.

The coefficients $f_{ix}({\bf q})$ and $f_{iy}({\bf q})$ are not independent
phenomenological relaxation constants. They arise from matrix elements of
the deformation-potential density operator between orthogonalized localized
orbitals. Their amplitudes and phases are correlated with the diagonal
form factors entering $f_{iz}({\bf q})$. For two DQDs coupled to the same
phonon modes, the scattering amplitudes of the individual dots interfere
\emph{before} the absolute square is taken in
Eq.~\eqref{eq:phonon_rate_microscopic}. Replacing
$|\langle n|F_1+F_2|m\rangle|^2$ by the sum of two independent probabilities
would discard the bath cross terms and must be justified by spatial
separation or an angular-average estimate. The parity-forbidden
$E_3\to E_0$ longitudinal channel at zero detuning is consequently a
statement about a particular projected operator, not a blanket prohibition
against acoustic decay. This distinction is consistent with microscopic
studies of orbital phonon coupling and of geometry-dependent relaxation in
silicon DQDs \cite{KrzywdaCywinski2021Phonons}.

The coherent parameters of Ref.~\cite{MacQuarrie2020} do not determine
wave-function extent, valley composition, or an acoustic-phonon-only
relaxation time. The reported charge-qubit scale
\(T_1<10\,\mathrm{ns}\) includes potentially nonphononic channels and
cannot be identified with \(1/\Gamma_{\rm ph}\); other silicon charge-qubit
regimes exhibit strongly different relaxation times
\cite{MacQuarrie2020,Wang2013ChargeRelax}.
The results below are therefore conditional on the explicitly stated
orbital geometry and the acoustic bath at \(100\,\mathrm{mK}\), assumed independent of classical detuning noise.

The control time of the first population-transfer maximum is of order
$0.27\,\mathrm{ns}$. The dimensionless parameter that tests whether the
phonon dissipator can visibly alter this peak is not $T_1$ alone but the
integrated escape rate $\int_0^{t_*}\!dt\,\Gamma_{\rm esc}(t)$,
where $\Gamma_{\rm esc}$ must be computed in the \emph{driven} basis.
In the weak-loss regime, this integral gives a leading-order estimate of the escape probability.
A weak correction during the pulse can coexist with substantial energy
loss during later storage. At the same time, the measured low-frequency
charge-noise amplitude does not validate a $1/f$ spectral continuation
up to the chosen $256\,\mathrm{GHz}$ cutoff. That continuation, especially
when used to derive relaxation rates near a DQD transition frequency,
should be regarded as a modeling hypothesis. Sensitivity to the high-frequency
rolloff and to the division between resolved colored fluctuations and
secular noise-induced transitions is part of the numerical validation, not
a posteriori graphical smoothing \cite{PaqueletWuetz2023,KrzywdaCywinski2021Phonons}.

\paragraph{Post-pulse storage dynamics.}
To distinguish driven charging from subsequent storage, let
\(t_*\) denote the first noiseless charging maximum, and define
\(\rho_{\rm ch}=U_0(t_*)\rho_\beta U_0^\dagger(t_*)\), where \(U_0\)
is generated by \(H_0(t)\).
At storage time \(\tau\geq0\), the tunnel drive is off and
\begin{equation}
\frac{d\rho_\alpha}{d\tau}
=-\frac{i}{\hbar}[H_B+H_n^{(\alpha)}(\tau),\rho_\alpha]
+\mathcal L_{\rm ph}[\rho_\alpha],\qquad
\rho_\alpha(0)=\rho_{\rm ch}.
\label{eq:storage_stochastic_gkls}
\end{equation}
The static-basis acoustic dissipator \(\mathcal L_{\rm ph}\) is obtained
from the microscopic finite-temperature phonon rates under weak-coupling
and secular approximations, accounting for transitions sharing Bohr
frequencies when they couple to common bath modes.
This is a \emph{post-pulse} model and does not claim phonon dissipation
has been included during the microwave pulse; a driven Floquet--Markov
treatment would be required for a complete charge--store cycle.

The quantum and classical parts of the noise model can now be combined
consistently. For each detuning-noise realization $\alpha$, the driven battery undergoes
unitary evolution from the same thermal state $\rho_\beta$ under
$H_\alpha(t)$, whereas the acoustic term in
Eq.~\eqref{eq:storage_stochastic_gkls} contributes only after the pulse.
The density matrix used for all observables is the ensemble average,
\begin{equation}
\rho(t)\equiv\overline{\rho}(t)
=\frac{1}{N_r}\sum_{\alpha=1}^{N_r}\rho_\alpha(t),
\label{eq:ensemble_average_state}
\end{equation}
with the analogous average applied to the storage states
$\rho_\alpha(\tau)$. In particular, ergotropy is calculated
\emph{from the averaged state}, rather than by averaging the ergotropies
of individual realizations. Details of the stochastic propagation and its numerical implementation are
given in Appendix~\ref{app:numerical_simulations}; the orbital factors,
phonon-rate integrals and spectral discretization are specified in
Appendix~\ref{app:phonon_repro_added}.

\subsection{Extractable work observables and ergotropy decomposition}
\label{subsec:ergotropy_decomposition}

All extractable work observables are evaluated from the noise-averaged state $\rho(t)$ with respect to the static battery Hamiltonian $H_B$. In this work, the central performance quantity is the ergotropy, which measures the part of the battery energy that can be extracted by a unitary operation. This choice is important for a noisy driven multilevel system because a large energy expectation value does not necessarily imply a large amount of useful work.
\begin{figure*}[t]
\centering
\includegraphics[width=0.98\textwidth]{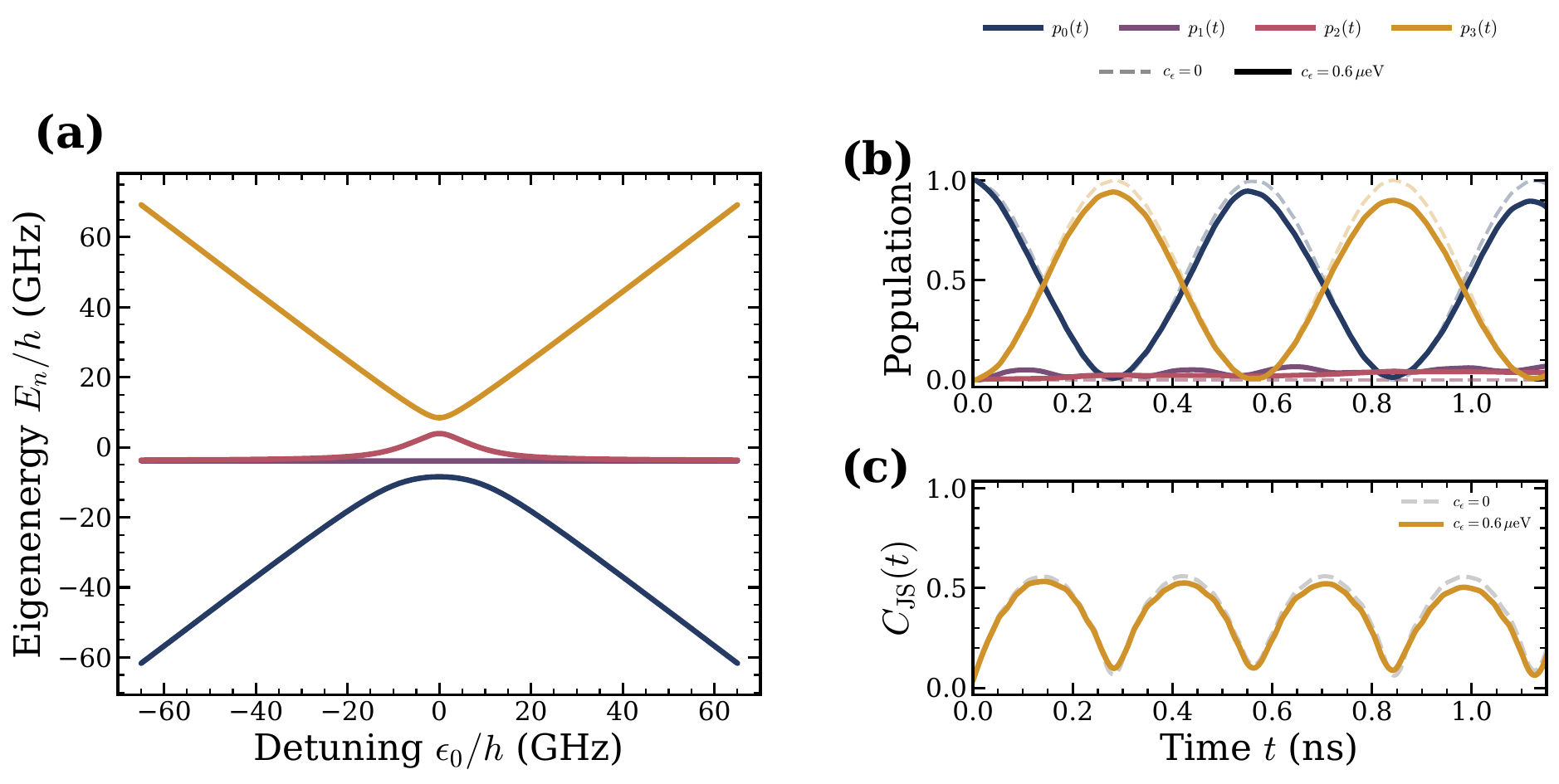}

\caption{Static spectrum and noise-averaged resonant microwave charging of the capacitively coupled double-quantum-dot battery. (a) Eigenenergies $E_n/h$ of the static Hamiltonian $H_B$ as a function of the common static detuning, $\epsilon_{0,1}/h=\epsilon_{0,2}/h\equiv\epsilon_0/h$. The spectrum is calculated for $\Delta_{0,1}/h=8.4\,\mathrm{GHz}$, $\Delta_{0,2}/h=6.6\,\mathrm{GHz}$, and $J/h=15.3\,\mathrm{GHz}$. At the symmetric operating point $\epsilon_0=0$, the four energy levels satisfy $E_n/h\simeq(-8.419,-3.929,3.929,8.419)\,\mathrm{GHz}$, giving the selected transition frequency $f_{03}=(E_3-E_0)/h\simeq16.838\,\mathrm{GHz}$. (b) Time-dependent energy-basis populations $p_n(t)=\langle\phi_n|\rho(t)|\phi_n\rangle$ for an initial Gibbs state at $T_{\mathrm{init}}=20\,\mathrm{mK}$, charged by symmetric tunnel-splitting modulation with $A_\Delta/h=4.0\,\mathrm{GHz}$ and $f_d=f_{03}$. The complementary evolution of $p_0$ and $p_3$ identifies the dominant $E_0\leftrightarrow E_3$ transfer, while $p_1$ and $p_2$ remain weakly populated. Dashed and solid traces denote noiseless and noise-averaged dynamics, respectively. (c) Jensen--Shannon coherence $C_{\mathrm{JS}}(t)$ in the same energy eigenbasis for $c_\epsilon=0$ and $0.6\,\mu\mathrm{eV}$. The $1/f$ detuning fluctuations are sampled over the assumed band $\omega_l/(2\pi)=1\,\mathrm{Hz}$ to $\omega_h/(2\pi)=256\,\mathrm{GHz}$. Ensemble averaging reduces the oscillation contrast and the energy-basis coherence without eliminating the selected charging pathway.}
\label{fig:spectrum_dynamics}
\end{figure*}

The total ergotropy is defined as \cite{binder2015quantum,campaioli2017enhancing}
\begin{equation}
    \xi(t)=\Tr\left[H_B\rho(t)\right]-\Tr\left[H_B\pi(t)\right],
    \label{eq:ergotropy}
\end{equation}
where $\pi(t)$ is the passive state associated with $\rho(t)$. The passive state is constructed by assigning the eigenvalues of $\rho(t)$, ordered from largest to smallest, to the eigenstates of $H_B$, ordered from lowest to highest energy, following the standard passivity construction \cite{PuszWoronowicz1978,Lenard1978}. Thus, $\xi(t)$ directly quantifies the useful work stored in the state after ensemble averaging over the detuning noise realizations.

To identify the physical origin of the extractable work, we separate the ergotropy into population and coherent contributions \cite{Francica2020}. We first dephase the state in the energy eigenbasis of $H_B$,
\begin{equation}
    \rho_d(t)=\sum_k \mel{\phi_k}{\rho(t)}{\phi_k}\ket{\phi_k}\bra{\phi_k}.
    \label{eq:dephased_state}
\end{equation}
The population ergotropy is then defined as the ergotropy of this dephased state,
\begin{equation}
    \xi_P(t)=\Tr\left[H_B\rho_d(t)\right]-\Tr\left[H_B\pi_d(t)\right],
    \label{eq:population_ergotropy}
\end{equation}
where $\pi_d(t)$ is the passive state associated with $\rho_d(t)$. The coherent contribution is obtained by subtraction,
\begin{equation}
    \xi_C(t)=\xi(t)-\xi_P(t).
    \label{eq:coherent_ergotropy}
\end{equation}
The population contribution $\xi_P(t)$ measures the extractable work stored in the diagonal energy-basis populations, namely in non-passive population orderings. By contrast, $\xi_C(t)$ measures the additional extractable work that is lost when the state is dephased in the energy-basis. Thus, $\xi_C(t)$ is associated with energy-basis coherences, whereas $\xi_P(t)$ is associated with non-passive population orderings, including but not limited to population inversion.
The population contribution has a direct passivity interpretation. In the energy eigenbasis of $H_B$, the diagonal populations are
\begin{equation}
    p_n(t)
    =
    \mel{\phi_n}{\rho(t)}{\phi_n},
    \qquad
    E_0\leq E_1\leq E_2\leq E_3 .
    \label{eq:energy_populations}
\end{equation}
Since $\rho_d(t)$ is diagonal in this basis, its ergotropy is entirely determined by the ordering of the populations. For the nondegenerate spectrum considered here, the dephased state $\rho_d(t)$ is passive if and only if
\begin{equation}
    p_0(t)\geq p_1(t)\geq p_2(t)\geq p_3(t).
    \label{eq:passive_ordering_condition}
\end{equation}
Consequently, \(
    \xi_P(t)>0
    \Longleftrightarrow 
    \rho_d(t)\ \text{is non-passive}.\)
Indeed, if the populations satisfy Eq.~\eqref{eq:passive_ordering_condition}, then $\rho_d(t)$ is already passive and $\pi_d(t)=\rho_d(t)$, so that $\xi_P(t)=0$. Conversely, if there exists a pair $n<m$ such that $p_m(t)>p_n(t)$, then the higher energy level $E_m$ is more populated than the lower energy level $E_n$. Exchanging these two populations decreases the energy by
\begin{equation}
    \Delta U_{nm}
    =
    (E_m-E_n)\,[p_m(t)-p_n(t)]>0,
    \label{eq:pairwise_energy_decrease}
\end{equation}
which shows that the diagonal state is non-passive and that a finite amount of work can be extracted from the population ordering. Therefore, the population ergotropy is positive exactly when the diagonal energy-basis populations violate passive ordering.

To visualize this mechanism, we introduce the passive ordering violation indicator
\begin{equation}
    \mathcal{V}_{\mathrm{pass}}(t)
    =
    \sum_{0\leq n<m\leq 3}
    (E_m-E_n)
    \left[p_m(t)-p_n(t)\right]_+ ,
    \label{eq:passive_violation_indicator}
\end{equation}
where $[x]_+=\max(x,0)$. This quantity is zero when the diagonal state is passive and becomes positive whenever at least one pair of populations violates the passive ordering. It is therefore a diagnostic of the same non-passivity condition that gives $\xi_P(t)>0$. In particular, it detects partial violations such as $p_3>p_1$ or $p_3>p_2$, even when the stronger highest-lowest inversion condition $p_3>p_0$ is not yet satisfied. However, $\mathcal{V}_{\mathrm{pass}}(t)$ should not be interpreted as an additional ergotropy contribution. In general, $\mathcal{V}_{\mathrm{pass}}(t)\neq\xi_P(t)$, because $\mathcal{V}_{\mathrm{pass}}$ sums all pairwise ordering violations, whereas $\xi_P$ is obtained from the globally passive rearrangement of $\rho_d(t)$. Thus, $\mathcal{V}_{\mathrm{pass}}$ is used only to identify and visualize the population ordering pathway responsible for the population ergotropy.

The role of coherence in the selected resonant transfer is assessed below by comparing $\xi_C$, $\xi_P$ and $C_{\mathrm{JS}}$ during the first charging branch.
\begin{figure*}[t]
\centering
\includegraphics[width=0.98\textwidth]{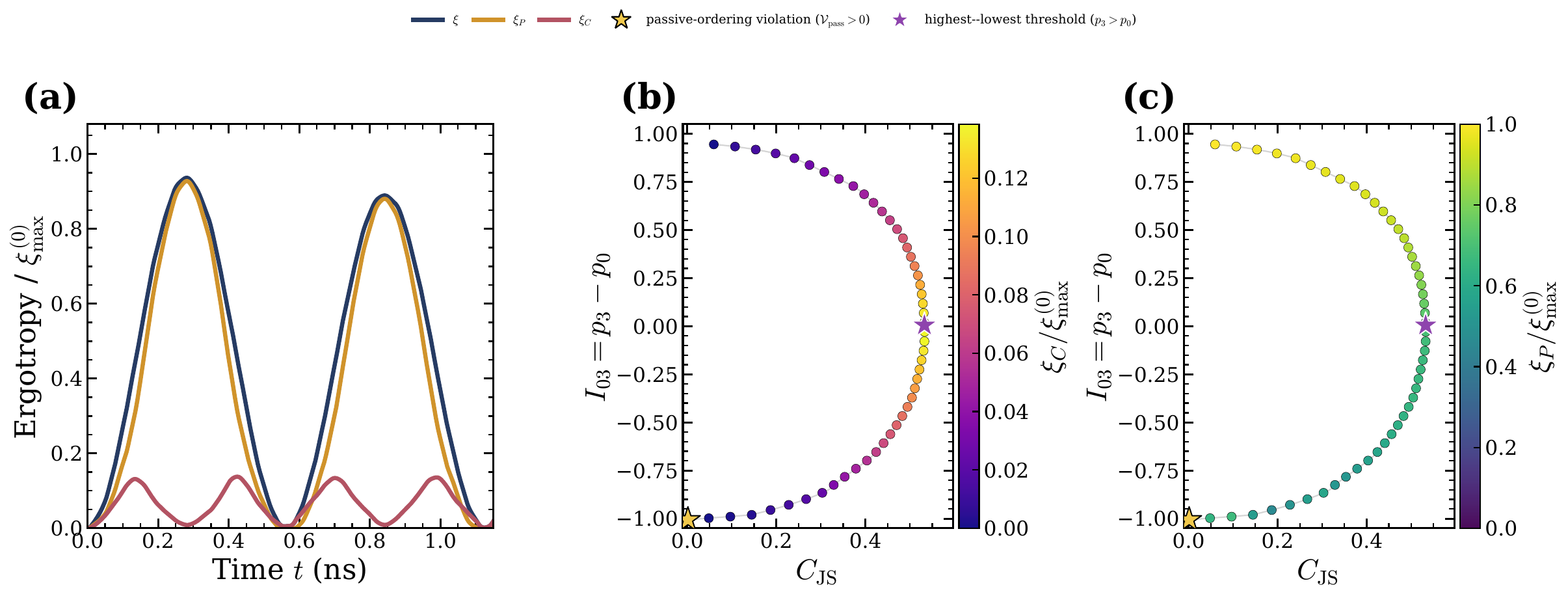}
\caption{Ergotropy decomposition and passive-ordering pathway during resonant charging under band-limited $1/f$ detuning noise. All panels use the static and driving parameters of Fig.~\ref{fig:spectrum_dynamics}, with a Gibbs preparation at $T_{\mathrm{init}}=20\,\mathrm{mK}$, a representative noise strength $c_\epsilon=0.6\,\mu\mathrm{eV}$, and cutoffs $\omega_l/(2\pi)=1\,\mathrm{Hz}$ and $\omega_h/(2\pi)=256\,\mathrm{GHz}$. (a) Total ergotropy $\xi(t)$, population ergotropy $\xi_P(t)$, and coherent ergotropy $\xi_C(t)$, each normalized by the maximum total ergotropy of the noiseless first charging cycle, $\xi_{\max}^{(0)}$. The close agreement of $\xi$ and $\xi_P$ near the principal charging maximum shows that the dominant extractable work is stored in the energy-basis population distribution. (b) First charging branch in the $(C_{\mathrm{JS}},I_{03})$ plane, where $I_{03}=p_3-p_0$ is the highest--lowest population imbalance. The trajectory is color-coded by $\xi_C/\xi_{\max}^{(0)}$. The markers distinguish a numerically resolved passive-ordering violation, $\mathcal V_{\mathrm{pass}}>0$, from the stronger condition $p_3>p_0$. At $20\,\mathrm{mK}$ the exact thermal non-passivity threshold is extremely small, so the plotted first marker depends on numerical resolution; the second marks the much later highest--lowest inversion. (c) The same trajectory, with marker colors showing the population fraction $\xi_P/\xi$. This ratio is displayed only where $\xi$ exceeds the numerical threshold, avoiding division by nearly vanishing ergotropy. The trajectories connect transient coherence with the formation of extractable work from non-passive populations.}
\label{fig:ergotropy_decomposition}
\end{figure*}

For the common charged reference state \(\rho_{\rm ch}\), the retained
population ergotropy during storage is quantified by
\begin{equation}
R_P(\tau,\kappa)=
\frac{\xi_P[\overline\rho_\kappa(\tau)]}{\xi_P[\rho_{\rm ch}]},
\qquad
\overline\rho_\kappa(\tau)=\frac1{N_r}\sum_{\alpha=1}^{N_r}
\rho_\alpha(\tau).
\label{eq:retention_new}
\end{equation}
Thus \(R_P\) is evaluated \emph{after} stochastic ensemble averaging,
with the same battery Hamiltonian and identical initial charged state
for every value of \(\kappa\).

\subsection{Jensen--Shannon coherence}
\label{subsec:js_coherence}

Quantum coherence is defined relative to the energy eigenbasis of $H_B$, which also determines passivity~\cite{Baumgratz2014,Streltsov2017}. Its incoherent reference is the dephased state $\rho_d(t)$ of Eq.~\eqref{eq:dephased_state}.

To quantify this coherence, we use the quantum Jensen--Shannon divergence, which provides a symmetric and bounded distinguishability measure between two quantum states \cite{Majtey2005,Lamberti2008,Radhakrishnan2016}. It is defined as
\begin{equation}
    J\left(\rho(t),\rho_d(t)\right)
    =
    S\left[\frac{\rho(t)+\rho_d(t)}{2}\right]
    -\frac{1}{2}S\left[\rho(t)\right]
    -\frac{1}{2}S\left[\rho_d(t)\right],
    \label{eq:js_divergence}
\end{equation}
where
\begin{equation}
    S(\rho)=-\Tr\!\left[\rho\log_2\rho\right]
    \label{eq:vonneumann_entropy}
\end{equation}
is the von Neumann entropy. The logarithm is taken in base 2 in the numerical calculations. The Jensen--Shannon coherence is then defined as
\begin{equation}
    C_{\mathrm{JS}}(t)=\sqrt{J\left(\rho(t),\rho_d(t)\right)}.
    \label{eq:js_coherence}
\end{equation}
The quantity $C_{\mathrm{JS}}(t)$ measures the distance between the noise-averaged battery state and its energy-dephased counterpart. It is particularly useful for the present problem because it remains well defined for mixed states, including the ensemble-averaged states generated by band-limited $1/f$ detuning noise. Importantly, $C_{\mathrm{JS}}(t)$ is not identified with the total ergotropy. Instead, it is analyzed together with $\xi(t)$, $\xi_P(t)$, $\xi_C(t)$, and $\mathcal{V}_{\mathrm{pass}}(t)$ in order to determine whether useful work is stored mainly in energy-basis coherence or in non-passive populations.
\begin{figure*}[t]
\centering
\includegraphics[width=0.98\textwidth]{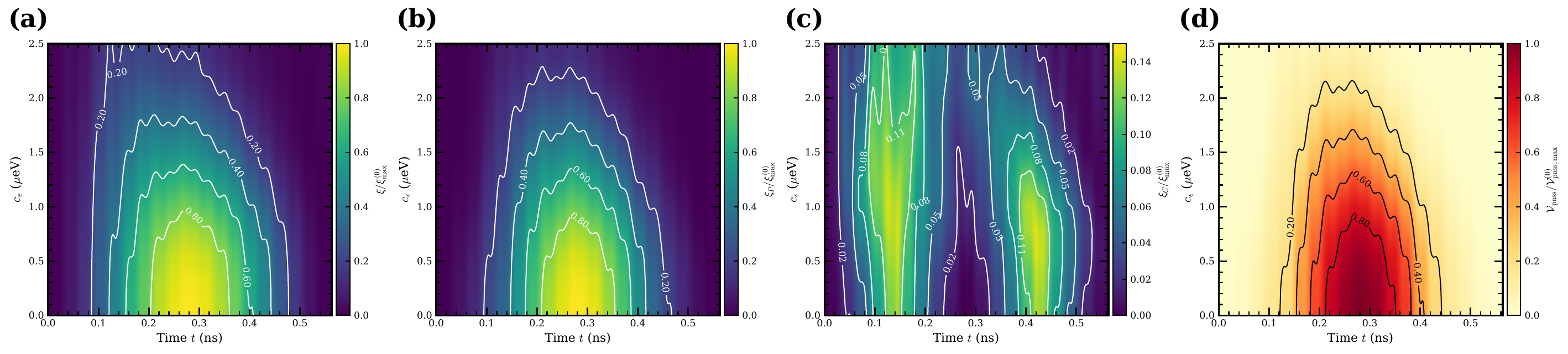}
\caption{Two-dimensional maps of extractable-work channels and passive-ordering violations during the first resonant charging branch. The Hamiltonian, symmetric microwave drive, and initial Gibbs temperature $T_{\mathrm{init}}=20\,\mathrm{mK}$ are the same as in Figs.~\ref{fig:spectrum_dynamics} and~\ref{fig:ergotropy_decomposition}. The local detuning fluctuations are independent ($\kappa=0$) and follow the assumed band-limited $1/f$ spectrum between $\omega_l/(2\pi)=1\,\mathrm{Hz}$ and $\omega_h/(2\pi)=256\,\mathrm{GHz}$; the spectral amplitude is scanned over $0\le c_\epsilon\le2.5\,\mu\mathrm{eV}$. (a) Normalized total ergotropy $\xi(t,c_\epsilon)/\xi_{\max}^{(0)}$. (b) Normalized population ergotropy $\xi_P(t,c_\epsilon)/\xi_{\max}^{(0)}$. (c) Normalized coherent ergotropy $\xi_C(t,c_\epsilon)/\xi_{\max}^{(0)}$, shown with an expanded color scale to resolve its smaller absolute magnitude. (d) Passive-ordering violation indicator $\mathcal V_{\mathrm{pass}}(t,c_\epsilon)/\mathcal V_{\mathrm{pass},\max}^{(0)}$, which is positive where the diagonal energy-basis populations cease to be passive; it is a diagnostic, not an additional ergotropy term. Here $\xi_{\max}^{(0)}$ and $\mathcal V_{\mathrm{pass},\max}^{(0)}$ are the respective maxima of the noiseless first charging branch. The similarity between (a), (b), and (d) links the reduction of useful work to the weakening of non-passive population transfer.}

\label{fig:ergotropy_maps}
\end{figure*}

\section{Results and discussion}
\label{sec:results}

\subsection{Spectral selection and noise-averaged coherent charging}
\label{subsec:spectral_selection}

Fig.~\ref{fig:spectrum_dynamics}(a) identifies the interacting spectrum at the operating point used for resonant charging. The energy gap $E_3-E_0$ and the parity-allowed tunnel-drive matrix element select the charging transition, while the intermediate states remain in the opposite parity sector. The complementary oscillations of $p_0(t)$ and $p_3(t)$ in Fig.~\ref{fig:spectrum_dynamics}(b) confirm that the microwave response is concentrated in this doublet. Under detuning noise, the oscillation contrast decreases without a qualitative change of the principal transfer pathway.

The Jensen--Shannon coherence in Fig.~\ref{fig:spectrum_dynamics}(c) supplies a complementary diagnostic. It is finite during the resonant exchange, demonstrating the formation of superpositions in the energy eigenbasis, but decreases after averaging over realizations with different detuning histories. This distinction is important: the suppression of the averaged coherence does not imply that individual stochastic trajectories undergo irreversible phonon relaxation during the charging stage. Their evolution remains unitary, and the loss of contrast follows from ensemble mixing.

\subsection{Population-dominated ergotropy and passive-ordering violation}
\label{subsec:population_dominated_ergotropy}
Fig.~\ref{fig:ergotropy_decomposition} relates the ergotropy decomposition to the sequence of passive-ordering changes along the first charging branch.
In Fig.~\ref{fig:ergotropy_decomposition}(a), $\xi(t)$ closely follows $\xi_P(t)$ near the principal charging maximum, whereas $\xi_C(t)$ is smaller. Finite $C_{\mathrm{JS}}$ in Fig.~\ref{fig:spectrum_dynamics}(c) nevertheless confirms that coherence accompanies the transfer: the distinction concerns how the accumulated extractable work is partitioned, not whether the drive is coherent.

The population-dominated maximum in Fig.~\ref{fig:ergotropy_decomposition}(a)
has an analytical explanation in the complete four-level Hilbert space.
For ideal ground-state charging confined to
\(\mathrm{span}\{\ket{\phi_0},\ket{\phi_3}\}\), Appendix~\ref{app:temperature_20_228}
gives \(\xi_C^{(0)}=\Delta_{10}\min\{q,1-q\}\), where
\(q\) is the transferred population and \(\Delta_{nm}=E_n-E_m\).
Consequently \(\xi_C^{(0)}/\Delta_{30}\leq0.1333\); the intermediate
energy \(E_1\) enters the passive rearrangement even if the drive does
not populate that level. This result explains the limited coherent
contribution in the nearly pure \(20\,\mathrm{mK}\) benchmark, without
assuming that coherence is absent during charging.
The finite-temperature extension, shown in Fig.~\ref{fig:thermal_onset_appendix},
introduces a nonzero transfer threshold
\(q_{\rm NP}(228\,\mathrm{mK})\simeq0.04679\) and reduces the
ideal maximum available work, while preserving population-dominated
storage near the first charge maximum. These analytical benchmarks do
not establish finite-temperature robustness under colored noise.

The parametric trajectories in Fig.~\ref{fig:ergotropy_decomposition}(b) and Fig.~\ref{fig:ergotropy_decomposition}(c) separate two population-ordering thresholds. The first marker in Fig.~\ref{fig:ergotropy_decomposition}(b) denotes the numerically resolved condition $\mathcal V_{\mathrm{pass}}>0$, whereas the second corresponds to $I_{03}=p_3-p_0>0$. These events are not equivalent. Non-passivity can arise when the population of an intermediate level is overtaken, even though the ground state is still more occupied than the highest-energy level. At $20\,\mathrm{mK}$, the ideal thermal onset occurs extremely early; the first marker visible in a discrete trajectory is therefore dependent on the numerical resolution, as discussed in Appendix~\ref{app:temperature_20_228}.

Fig.~\ref{fig:ergotropy_decomposition}(c) follows the same trajectory with the population fraction $\xi_P/\xi$ as its color coordinate. Its increase identifies the part of the transfer during which non-passive populations carry most of the extractable work. The coherence recorded by $C_{\mathrm{JS}}$ remains meaningful: a noncommuting drive can generate superpositions and transfer energy~\cite{Caravelli2021}, yet these superpositions need not provide the largest contribution to ergotropy. The present balance between $\xi_C$ and $\xi_P$ reflects the selectivity of the $E_0\leftrightarrow E_3$ protocol and need not extend to drives involving substantial occupation of $E_1$ and $E_2$.

\subsection{Noise-induced weakening of the passive-ordering pathway}
\label{subsec:noise_population_inversion}

Fig.~\ref{fig:ergotropy_maps} displays the time and noise-strength dependence of $\xi$, $\xi_P$, $\xi_C$ and $\mathcal V_{\mathrm{pass}}$. The last quantity identifies non-passive population orderings but is not an additional extractable-work contribution.
\begin{figure*}[t]
\centering
\includegraphics[width=0.98\textwidth]{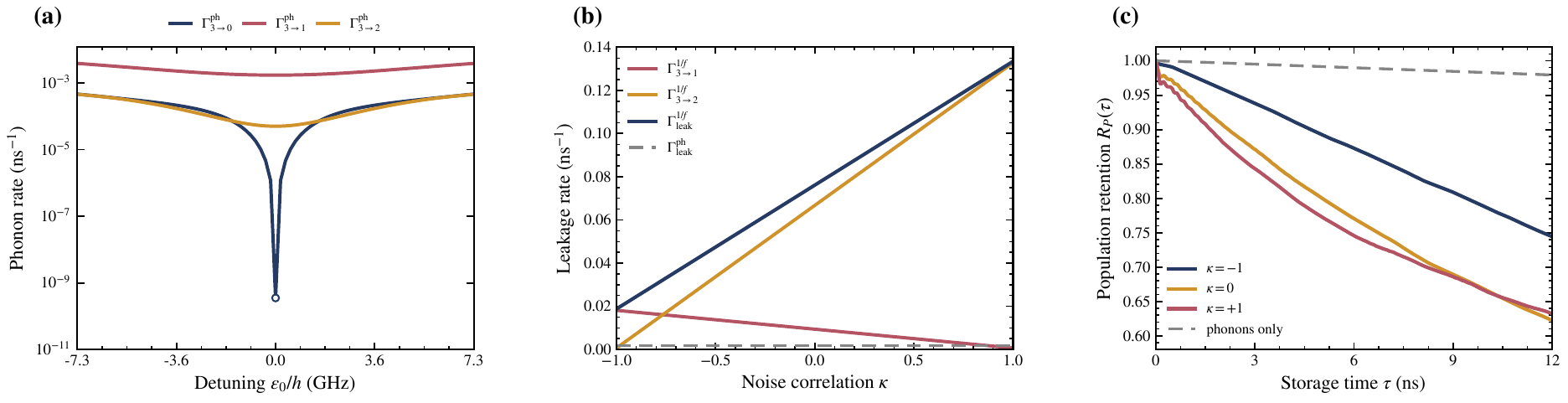}
\caption{Parity-resolved acoustic relaxation, correlated electrical leakage, and post-pulse population-ergotropy retention in the Si/SiGe coupled-DQD battery. The coherent Hamiltonian parameters are $\Delta_{0,1}/h=8.4\,\mathrm{GHz}$, $\Delta_{0,2}/h=6.6\,\mathrm{GHz}$, and $J/h=15.3\,\mathrm{GHz}$. The common charged reference state $\rho_{\mathrm{ch}}$ is prepared from a Gibbs state at $T_{\mathrm{init}}=20\,\mathrm{mK}$ by an ideal resonant tunnel-modulation pulse with $A_\Delta/h=4.0\,\mathrm{GHz}$ and $f_d=f_{03}\simeq16.838\,\mathrm{GHz}$; the microwave drive is switched off before the storage interval. (a) Calculated downward acoustic rates $\Gamma^{\mathrm{ph}}_{3\to n}$ ($n=0,1,2$) versus the common static detuning $\epsilon_0/h$ (in $\mathrm{GHz}$), for a phonon bath statistically independent of the classical detuning noise at $T_{\mathrm{ph}}=100\,\mathrm{mK}$ and illustrative orthogonalized Gaussian orbitals with $d=100\,\mathrm{nm}$, $r_{xy}=20\,\mathrm{nm}$, $r_z=2.5\,\mathrm{nm}$, and $D_c=200\,\mathrm{nm}$. The detuning interval in panel (a) is expressed as energy divided by $h$, with $1\,\mu\mathrm{eV}/h=0.241799\,\mathrm{GHz}$. At zero detuning, parity suppresses the longitudinal $3\to0$ matrix element, although the transverse orbital contribution need not vanish. (b) Weak-noise transition diagnostics $\Gamma^{1/f}_{3\to1}$, $\Gamma^{1/f}_{3\to2}$, and their sum as functions of the electrical cross-correlation coefficient $-1\le\kappa\le1$, compared with the acoustic escape-rate reference. The opposite $\kappa$ dependence of the two electrical channels reflects interference of their detuning matrix elements; these perturbative rates are not added to the sampled stochastic dynamics. (c) Population-ergotropy retention $R_P(\tau,\kappa)=\xi_P[\overline\rho_\kappa(\tau)]/\xi_P[\rho_{\mathrm{ch}}]$ during $0\le\tau\le12\,\mathrm{ns}$ for $\kappa=-1,0,+1$, including correlated classical detuning noise and acoustic relaxation, together with a phonon-only reference. The assumed noise strength is $c_\epsilon=0.6\,\mu\mathrm{eV}$ over $1\,\mathrm{Hz}$--$256\,\mathrm{GHz}$; the resulting comparison is model-specific and is not a device lifetime measurement.}
\label{fig:phonon_experimental_audit}
\end{figure*}

The region of high total ergotropy in Fig.~\ref{fig:ergotropy_maps}(a) closely follows the high-$\xi_P$ region in Fig.~\ref{fig:ergotropy_maps}(b), both concentrated around the first resonant population transfer. Fig.~\ref{fig:ergotropy_maps}(c) displays $\xi_C$ on an expanded color scale to resolve its smaller values. Reading Fig.~\ref{fig:ergotropy_maps}(c) on its own would therefore exaggerate the apparent contribution of coherence to the useful-work budget. The comparison of absolute scales establishes that the population channel dominates, although the coherent component remains finite along the charging branch.

Fig.~\ref{fig:ergotropy_maps}(d) provides the corresponding test in terms of population ordering. The indicator $\mathcal V_{\mathrm{pass}}$ is constructed from the same averaged populations but does not equal $\xi_P$ in magnitude. Its nonzero domain nevertheless identifies where the dephased state has become non-passive. As the spectral amplitude $c_\epsilon$ increases, both that domain and the high-$\xi_P$ region shrink. The agreement between Fig.~\ref{fig:ergotropy_maps}(b) and Fig.~\ref{fig:ergotropy_maps}(d) therefore supports the interpretation that detuning noise reduces useful work mainly by disrupting the formation of the non-passive population distribution. The coherent ergotropy in Fig.~\ref{fig:ergotropy_maps}(c) instead exhibits a more structured, nonmonotonic dependence on noise strength; its localized features do not offset the dominant reduction in population ergotropy near the charging maximum.
The reduced population-transfer contrast is consistent with realizations acquiring different detuning and effective resonance shifts during the pulse, as in inhomogeneously broadened semiconductor charge-qubit dynamics~\cite{Connors2019}. This picture also explains why the map of the coherent contribution need not mirror the map of total work: coherent amplitudes and population ordering respond differently to ensemble mixing. An improvement in the local electrostatic environment would be expected to favor both the driven transfer and its non-passive populations. Progress in lowering low-frequency charge noise in optimized $^{28}\mathrm{Si}/\mathrm{SiGe}$ structures provides an experimental motivation for such improvements~\cite{PaqueletWuetz2023}, but does not independently calibrate the high-frequency spectrum assumed here.
\subsection{Parity-resolved leakage and post-pulse ergotropy retention}
\label{sec:phonon_results_discussion}

Fig.~\ref{fig:phonon_experimental_audit} examines whether the ideally charged state retains its population ergotropy after the microwave pulse, separating acoustic relaxation, correlation-dependent electrical leakage and the resulting post-pulse dynamics.
The symmetry derived in Eq.~\eqref{eq:parity_exact} eliminates the longitudinal $E_3\to E_0$ single-phonon matrix element at zero detuning. The intermediate levels, which belong to the opposite parity sector, remain accessible through the longitudinal coupling. A small nonzero direct $3\to0$ rate is nevertheless allowed by the transverse projected orbital vertex; its magnitude is controlled by the overlap of localized wavefunctions rather than the parity selection rule alone. This accounts for the residual channel in Fig.~\ref{fig:phonon_experimental_audit}(a) without implying immunity to acoustic relaxation. Away from the symmetric point, even the longitudinal selection rule ceases to be exact.

The collective and differential detuning operators in Eq.~\eqref{eq:common_diff_new} couple the charged state to the intermediate levels with distinct interference signs. For real eigenvectors, the relevant matrix elements are $Z_{13}^{(1)}=0.6172$, $Z_{13}^{(2)}=-0.4205$, $Z_{23}^{(1)}=0.7868$, and $Z_{23}^{(2)}=0.9073$. Their opposite interference terms, together with the larger $1/f$ spectral weight at $f_{32}\simeq4.49\,\mathrm{GHz}$ than at $f_{31}\simeq12.35\,\mathrm{GHz}$, account for the correlation dependence of the total electrical leakage diagnostic in Fig.~\ref{fig:phonon_experimental_audit}(b) through Eq.~\eqref{eq:electrical_leak_exact}. These are bidirectional weak-noise diagnostics, not additional jump rates imposed on the sampled classical-noise trajectories.

The acoustic reference line in Fig.~\ref{fig:phonon_experimental_audit}(b) does not vary with $\kappa$ because the phonon bath and orbital geometry are held fixed. It is not a zero-dissipation reference. Similarly, the electrical curves are perturbative diagnostics of selected transitions rather than a complete prediction for extractable work. The distinction matters because detuning fluctuations do not commute with $H_B$ and can change both populations and energy expectation values. Fig.~\ref{fig:phonon_experimental_audit}(c) evaluates the nonlinear retention $R_P$ from the full ensemble-averaged density matrix, using the same initial state $\rho_{\mathrm{ch}}$ for every $\kappa$ [Eqs.~\eqref{eq:storage_stochastic_gkls} and~\eqref{eq:retention_new}].

For the chosen geometry, $s=\exp[-d^2/(4r_{xy}^2)]\simeq1.93\times10^{-3}$ suppresses the transverse orbital contribution, while the longitudinal form factor retains its dipolar dependence. At zero detuning the calculated acoustic rates are $\Gamma^{\rm ph}_{3\to0}=3.51\times10^{-10}$, $\Gamma^{\rm ph}_{3\to1}=1.72\times10^{-3}$, and $\Gamma^{\rm ph}_{3\to2}=5.03\times10^{-5}\,\mathrm{ns}^{-1}$. Their inverse summed escape rate is approximately $564\,\mathrm{ns}$, consistent with $R_P(12\,\mathrm{ns})\simeq0.979$ for phonons alone. This geometry-dependent quantity is neither a measured charge-qubit $T_1$ nor a universal ergotropy lifetime. The distinction from adiabatic electron transfer, where relaxation may facilitate charge motion, is relevant to the comparison with Krzywda and Cywi\'nski~\cite{KrzywdaCywinski2021Phonons}.

Within the specified $1/f$ model, Fig.~\ref{fig:phonon_experimental_audit}(c) shows greater population-ergotropy retention for anticorrelated fluctuations ($\kappa=-1$) than for independent or positively correlated noise. Fig.~\ref{fig:phonon_experimental_audit}(b) clarifies the competing mechanisms: anticorrelation reduces the $3\to2$ electrical leakage diagnostic while enhancing the $3\to1$ channel, which is weaker in the uncorrelated reference case. The full retention, however, follows from the nonlinear ergotropy of the ensemble-averaged density matrix, not from a single transition rate. The phonon-only reference remains more favorable over the displayed interval. Thus the advantage of anticorrelation is relative to the other noisy scenarios, under the assumed geometry and noise spectrum; it does not establish a universally protected storage lifetime.

The comparison concerns an ideal charged state followed by storage with a $100\,\mathrm{mK}$ phonon bath; it is not a fully dissipative charging cycle. The constant $\kappa$, high-frequency $1/f$ continuation and illustrative orbital geometry are modeling assumptions that require device-specific validation.
\par

\section{Conclusion}
\label{sec:conclusion}

In this work, we have investigated resonant charging and post-pulse work retention in a quantum battery formed by two capacitively coupled Si/SiGe double quantum dots. The charging protocol relies on microwave tunnel-coupling modulation under time-dependent $1/f$ detuning fluctuations. At the symmetric operating point, an exact parity symmetry of the interacting Hamiltonian selects the dominant transition between the lowest and highest energy eigenstates. The noise-averaged charging dynamics shows strong population redistribution accompanied by transient energy-basis coherence. By decomposing the total ergotropy into population and coherent contributions, we find that the dominant extractable work is stored in non-passive energy-level populations. The passive-ordering analysis also establishes that population ergotropy can appear before the stronger highest--lowest population-inversion threshold is reached. Detuning noise reduces the transfer contrast and weakens the population-ordering pathway responsible for useful work storage. The consistent behavior of the total ergotropy, population ergotropy, and passive ordering maps supports this interpretation for the selected charging protocol.

The post-pulse analysis identifies the environmental channels that affect the population ergotropy after the microwave field has been switched off. The parity symmetry suppresses the direct $E_3\rightarrow E_0$ longitudinal single-phonon matrix element at the symmetric point, while transitions involving the intermediate levels $E_1$ and $E_2$ remain allowed. The microscopic deformation-potential calculation shows that the remaining acoustic relaxation depends on the projected orbital form factors and the assumed dot geometry. Spatially correlated detuning fluctuations have an additional effect through interference between the two local transition amplitudes. In particular, the $E_3\rightarrow E_1$ and $E_3\rightarrow E_2$ electrical leakage diagnostics respond oppositely to the noise-correlation coefficient. For the specified model, the resulting storage dynamics shows greater population-ergotropy retention under anticorrelated fluctuations than under independent or positively correlated fluctuations. This links the symmetry of the charging transition and the structure of the environmental leakage channels to the useful work retained by the battery.
The quantitative retention predictions are conditional on the adopted orbital geometry, the assumed electrical cross-spectrum, and the extension of the $1/f$ model to high frequencies. The acoustic dissipator acts only during storage, starting from a common ideally charged state, so the calculation does not represent a fully dissipative charge--storage cycle. The complementary finite-temperature benchmark shows that thermal occupation reduces the available extractable work and changes the onset of population non-passivity, without establishing temperature-independent robustness under colored noise. These results provide a basis for examining how resonant charge control and environmental correlations affect useful-work storage in semiconductor quantum batteries. Device-specific orbital characterization, measured noise spectra, and a dissipative treatment during driven charging will be important for extending the present predictions to experimental implementations.

\begin{acknowledgments}
K. Loukhssami gratefully acknowledges financial support
from the National Center for Scientific and Technical
Research (CNRST, Morocco) through the PhD Associate
Scholarship PASS initiative. The authors also acknowledge
CNRST for providing access to the HPC MARWAN infrastructure
used for the numerical simulations reported in this work.
\end{acknowledgments}
\appendix
\section{Details of the numerical simulations}
\label{app:numerical_simulations}

The stochastic simulations use the band-limited $1/f$
detuning noise defined in Eq.~\eqref{eq:one_over_f_noise},
with $\kappa=0$ for
Figs.~\ref{fig:spectrum_dynamics}--\ref{fig:ergotropy_maps}.
For each noise realization $\alpha$, the density matrix
evolves unitarily according to
\begin{equation}
\rho_\alpha(t)=U_\alpha(t)\rho_\beta
U_\alpha^\dagger(t).
\label{eq:appendix_density_thermal}
\end{equation}
The ensemble-averaged state is
\begin{equation}
\overline{\rho}(t)=\frac{1}{N_r}
\sum_{\alpha=1}^{N_r}\rho_\alpha(t).
\label{eq:appendix_average_gibbs}
\end{equation}
All observables are evaluated from $\overline{\rho}(t)$,
with $N_r=2\times10^4$ stochastic realizations.
For Fig.~\ref{fig:phonon_experimental_audit},
the storage dynamics starts from the same noiselessly
charged state for all noise-correlation scenarios.
The correlated detuning fluctuations follow
Eq.~\eqref{eq:noise_matrix_new}, while acoustic
relaxation is included only during the storage stage.
\par

\section{Analytical Gibbs populations and temperature distinction}
\label{app:temperature_20_228}
\label{app:temperature_robustness}
\setcounter{figure}{0}
\renewcommand{\thefigure}{B\arabic{figure}}
\renewcommand{\theHfigure}{appendixB.\arabic{figure}}
The equilibrium reference used in Figs.~\ref{fig:spectrum_dynamics}--\ref{fig:phonon_experimental_audit} is the Gibbs state at
$T_{\mathrm{init}}=20\,\mathrm{mK}$. Its energy-basis populations are
\begin{equation}
 p_n^\beta(T)=\frac{\exp[-(E_n-E_0)/(k_BT)]}
 {\sum_{m=0}^{3}\exp[-(E_m-E_0)/(k_BT)]}.
 \label{eq:app_boltzmann_pops}
\end{equation}
For $\Delta_{0,1}/h=8.4$, $\Delta_{0,2}/h=6.6$ and
$J/h=15.3\,\mathrm{GHz}$, $(E_1-E_0)/k_B\simeq215\,\mathrm{mK}$.
At $20\,\mathrm{mK}$, $p_0^\beta\simeq0.999979$ and
$p_1^\beta\simeq2.1\times10^{-5}$. At the electronic-temperature
benchmark $T_e=228\,\mathrm{mK}$ reported for the device motivating the
Hamiltonian parameters~\cite{MacQuarrie2020}, the ideal thermal weights are
\begin{equation}
\begin{aligned}
p_0^\beta&\simeq0.670294, &p_1^\beta&\simeq0.260521,\\
p_2^\beta&\simeq0.049821, &p_3^\beta&\simeq0.019364.
\end{aligned}
\label{eq:app_228_populations}
\end{equation}
The experimental pulsed initialization need not coincide with this Gibbs
preparation. The acoustic-bath temperature $T_{\mathrm{ph}}=100\,\mathrm{mK}$
in Fig.~\ref{fig:phonon_experimental_audit} is independent of $T_e$.

\begin{figure*}[!t]
\centering
\includegraphics[width=0.985\textwidth]{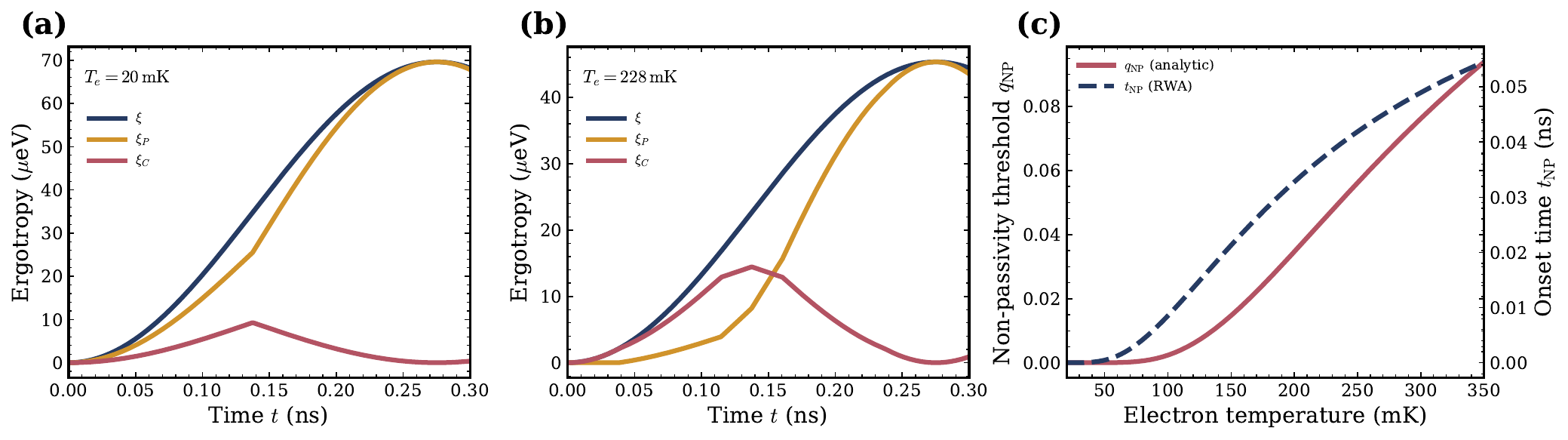}
\caption{Thermal Gibbs preparation, ergotropy decomposition, and the analytical onset of population non-passivity for ideal resonant charging of the coupled-DQD battery. All panels use $\Delta_{0,1}/h=8.4\,\mathrm{GHz}$, $\Delta_{0,2}/h=6.6\,\mathrm{GHz}$, $J/h=15.3\,\mathrm{GHz}$, and a symmetric tunnel drive with $A_\Delta/h=4.0\,\mathrm{GHz}$ at $f_d=f_{03}$. (a) Total ($\xi$), population ($\xi_P$), and coherent ($\xi_C$) ergotropies versus charging time for a Gibbs initial state at $T_e=20\,\mathrm{mK}$, where the near-ground-state preparation permits an almost complete first resonant transfer. (b) The same quantities at $T_e=228\,\mathrm{mK}$, illustrating the reduced amount of extractable work and the finite thermal threshold for population non-passivity, while the population contribution remains dominant near the charging maximum. (c) Non-passivity threshold $q_{\mathrm{NP}}(T_e)$ (left axis) obtained from the ordering of the initial Gibbs populations, and its corresponding onset time $t_{\mathrm{NP}}^{\mathrm{RWA}}(T_e)$ (right axis), calculated within the resonant rotating-wave approximation. At $228\,\mathrm{mK}$, $q_{\mathrm{NP}}\simeq0.04679$ and $t_{\mathrm{NP}}^{\mathrm{RWA}}\simeq0.03819\,\mathrm{ns}$. The curves in (a) and (b) are ideal RWA envelopes; counter-rotating micromotion of the full microwave Hamiltonian is not resolved. No charge noise or phonon dissipation is included, and the Gibbs preparation temperature $T_e$ is distinct from the $100\,\mathrm{mK}$ acoustic-bath temperature used for Fig.~\ref{fig:phonon_experimental_audit}.}
\label{fig:thermal_onset_appendix}
\end{figure*}

\paragraph{Exact pure-state benchmark.}
Because $[H_B,\Pi]=[D,\Pi]=0$, ideal symmetric charging preserves parity.
An initial ground state evolves within
$\operatorname{span}\{\ket{\phi_0},\ket{\phi_3}\}$; write
$\ket{\psi(t)}=a(t)\ket{\phi_0}+b(t)\ket{\phi_3}$ and
$p=|b|^2$. Its diagonal populations are $(1-p,0,0,p)$.
With $\Delta_{nm}=E_n-E_m$, passive rearrangement in the \emph{full
four-level battery}, rather than in an artificially truncated two-level
spectrum, yields
\begin{align}
\xi^{(0)}&=p\Delta_{30},\nonumber\\
\xi_P^{(0)}&=\begin{cases}
p\Delta_{31},&0\le p\le\frac12,\\
p\Delta_{30}-(1-p)\Delta_{10},&\frac12\le p\le1,
\end{cases}\nonumber\\
\xi_C^{(0)}&=\Delta_{10}\min\{p,1-p\}.
\label{eq:ideal_ergotropy_exact}
\end{align}
The $E_1$ level enters the passive rearrangement even though it is
unoccupied during ideal ground-state charging. This gives the spectral
coherent-ergotropy ceiling
\begin{equation}
 \frac{\xi_C^{(0)}}{\Delta_{30}}\le
 \frac{\Delta_{10}}{2\Delta_{30}}\simeq0.1333,
 \label{eq:ideal_coherence_ceiling}
\end{equation}
with equality for $p=1/2$. It explains the scale of the coherent peak
in Fig.~\ref{fig:ergotropy_decomposition}(a), but does not constitute
an exact identity for a noisy ensemble-averaged mixed state.

\paragraph{Finite-temperature passive ordering.}
This appendix also tests the passive-ordering interpretation without
introducing an additional environmental channel. Define
$d_\beta=p_0^\beta-p_3^\beta>0$ and the transition probability
$q(t)=|\mel{\phi_3}{U_0(t)}{\phi_0}|^2$, where $U_0(t)$ is the
unitary propagator of $H_B+H_{\mathrm{ch}}(t)$.
The odd-parity Gibbs populations remain unchanged under the
symmetric ideal drive, while the two even-parity populations satisfy
\begin{align}
p_0(t)&=p_0^\beta-d_\beta q(t),&
p_1(t)&=p_1^\beta,\nonumber\\
p_2(t)&=p_2^\beta,&
p_3(t)&=p_3^\beta+d_\beta q(t).
\label{eq:app_mixed_populations}
\end{align}
Since $\rho_\beta$ is initially passive and a unitary operation
preserves its spectrum, the \emph{total} ergotropy is exactly
\begin{equation}
\xi_{\mathrm{ideal}}(t,T_e)=d_\beta q(t)(E_3-E_0).
\label{eq:app_thermal_total}
\end{equation}
The population contribution is obtained by sorting the four diagonal
weights $p_n(t)$ in nonincreasing order. Writing these ordered weights
as $p_n^{\downarrow}(t)$ gives the globally valid result
\begin{align}
\xi_P(t)&=\sum_{n=0}^3 E_n
\bigl[p_n(t)-p_n^{\downarrow}(t)\bigr],\nonumber\\
\xi_C(t)&=\xi_{\mathrm{ideal}}(t,T_e)-\xi_P(t).
\label{eq:app_thermal_split}
\end{align}
The initial Gibbs ordering is
$p_0^\beta>p_1^\beta>p_2^\beta>p_3^\beta$.
The \emph{first} violation occurs when either $p_3>p_2$
or $p_0<p_1$, so that
\begin{equation}
q_{\mathrm{NP}}(T_e)=\min\!\left\{
\frac{p_2^\beta-p_3^\beta}{d_\beta},
\frac{p_0^\beta-p_1^\beta}{d_\beta}
\right\}.
\label{eq:app_q_np}
\end{equation}
At $228\,\mathrm{mK}$, these two candidates equal
$0.0467903$ and $0.6295185$, respectively. The first onset is thus
$p_3=p_2$ at $q_{\mathrm{NP}}\simeq0.0468$; it precedes the later
highest--lowest population inversion. The non-passivity onset is also
distinct from the later crossover at which $\xi_P$ may exceed $\xi_C$.
In the ideal resonant rotating-wave approximation (RWA),
\begin{align}
\Omega_{03}^{\mathrm{RWA}}
 &=\frac{A_\Delta |M_{30}^{(\Delta)}|}{2\hbar},
 &q_{\mathrm{RWA}}(t)&=\sin^2\!\left(\frac{\Omega_{03}^{\mathrm{RWA}}t}{2}\right),
\label{eq:app_omega_rwa}
\end{align}
and the estimated onset time is
\begin{equation}
 t_{\mathrm{NP}}^{\mathrm{RWA}}(T_e)
 =\frac{2}{\Omega_{03}^{\mathrm{RWA}}}
 \arcsin\!\sqrt{q_{\mathrm{NP}}(T_e)}.
 \label{eq:app_t_np}
\end{equation}
For $A_\Delta/h=4.0\,\mathrm{GHz}$, one finds
$\Omega_{03}^{\mathrm{RWA}}\simeq11.4185\,\mathrm{ns}^{-1}$ and
$t_{\mathrm{NP}}^{\mathrm{RWA}}(228\,\mathrm{mK})\simeq
0.03819\,\mathrm{ns}$. This onset estimate is based on RWA; the
non-RWA pulse dynamics need not have an identical crossing time.

Fig.~\ref{fig:thermal_onset_appendix}(a) and Fig.~\ref{fig:thermal_onset_appendix}(b)
show the smooth RWA-envelope benchmark and illustrate how the finite thermal occupation of the intermediate states
delays the appearance of non-passive diagonal populations, even though
the resonant transfer remains parity selected. The increasing threshold
in Fig.~\ref{fig:thermal_onset_appendix}(c) follows from the required redistribution needed to overcome
the initial Gibbs ordering; it is not evidence for an increase of
quantum coherence with temperature. At higher $T_e$ the total available
work decreases, while the threshold for $\xi_P>0$ is shifted to a
later point in the transfer. These ideal calculations do not establish
thermal robustness under the colored detuning noise of Figs.~\ref{fig:spectrum_dynamics}--\ref{fig:phonon_experimental_audit}.

\section{Orbital factors and numerical checks for the post-pulse test}
\label{app:phonon_repro_added}
For each DQD we use normalized localized envelopes
$\psi_a(\mathbf r)\propto\exp[-(x-x_a)^2/(2r_{xy}^2)
-y^2/(2r_{xy}^2)-z^2/(2r_z^2)]$,
with centers $x_{i,L/R}=X_i\mp d/2$, $X_{1/2}=\mp D_c/2$.
The two nonorthogonal orbitals in each pair have overlap
$s=\exp[-d^2/(4r_{xy}^2)]$. A symmetric L\"owdin transformation,
$B=S^{-1/2}$ with
$S=\bigl(\begin{smallmatrix}1&s\\s&1\end{smallmatrix}\bigr)$,
gives orthonormalized form factors $F^{\mathrm{orth}}=BF^{(0)}B$.
Before orthogonalization the plane-wave-density matrix elements are
\begin{widetext}
\begin{equation}
F^{(0)}_{ab}(\mathbf q)=
\exp\!\left[-\frac{r_{xy}^2(q_x^2+q_y^2)+r_z^2q_z^2}{4}
-\frac{(x_a-x_b)^2}{4r_{xy}^2}
+\frac{iq_x(x_a+x_b)}{2}\right].
\end{equation}
\end{widetext}
For a phonon frequency $\omega_{mn}=(E_m-E_n)/\hbar$, angular
integration of Eq.~\eqref{eq:phonon_rate_microscopic} yields
\begin{widetext}
\begin{equation}
\Gamma_{m\to n}^{\mathrm{ph}}=
\sum_{\lambda}\frac{q_\lambda^3}{8\pi^2\rho_m\hbar v_\lambda^2}
[n_B(\omega_{mn},T_{\mathrm{ph}})+1]
\int d\Omega\;|\mathcal D_\lambda(\hat{\bf q})|^2
\left|\mel{\phi_n}{\sum_i F_i^{\mathrm{orth}}(\mathbf q)}{\phi_m}\right|^2,
\label{eq:phonon_integral_explicit_added}
\end{equation}
\end{widetext}
where $q_\lambda=\omega_{mn}/v_\lambda$ and
$\mathcal D_\lambda=\Xi_d(\hat{\bf q}\cdot\mathbf e_\lambda)
+\Xi_u(\hat{\bf q}\cdot\hat{\bf z})(\mathbf e_\lambda\cdot\hat{\bf z})$.
The evaluation supplied with the code uses a 24-point Gauss--Legendre
quadrature for $\cos\theta$ and a 48-point equidistant azimuthal quadrature.
For the stated illustrative geometry at zero detuning the numerical
acoustic downward rates are
$\Gamma_{3\to0}^{\rm ph}\simeq3.51\times10^{-10}\,\mathrm{ns^{-1}}$,
$\Gamma_{3\to1}^{\rm ph}\simeq1.72\times10^{-3}\,\mathrm{ns^{-1}}$,
and $\Gamma_{3\to2}^{\rm ph}\simeq5.03\times10^{-5}\,\mathrm{ns^{-1}}$.
Their quantitative device relevance cannot be inferred without measured
orbital geometry and a verified acoustic coupling model.

For the colored-noise implementation, a finite record of duration $T$
resolves frequencies $f_k=k/T$ above $1/T$ and includes the portion
$1\,\mathrm{Hz}<f<1/T$ as a Gaussian quasi-static Fourier-band component
with variance $2c_\epsilon^2\ln[(1/T)/(1\,\mathrm{Hz})]$.
This is part of the same spectrum, not added disorder. A sampling interval
of $1\,\mathrm{ps}$ provides a Nyquist frequency of $500\,\mathrm{GHz}$,
above the assumed $256\,\mathrm{GHz}$ cutoff.
The numerical workflow evaluates the parity matrix elements,
acoustic transition rates, and correlation-resolved electrical-rate
diagnostics separately from the post-pulse stochastic master-equation propagation in Lindblad form.
The population ergotropy during storage is evaluated from the
ensemble-averaged state, consistently with Eq.~\eqref{eq:retention_new}.
The stochastic detuning model retains the prescribed spectral band for
each value of $\kappa$; the perturbative electrical-rate diagnostics of
Fig.~\ref{fig:phonon_experimental_audit}(b) are not added a second time
to this propagation. The adopted $1/f$ spectrum is extrapolated to
GHz-scale Bohr frequencies, so extending these predictions to a specific
device requires independent information about its high-frequency noise
and orbital geometry. The discretization and quadrature settings used for the reported calculations are stated above; the stochastic ensemble sizes and reference integration steps are specified in Appendix~\ref{app:numerical_simulations}.

\section*{Data Availability}
The numerical data and simulation scripts supporting the results of this study are available from the corresponding author upon reasonable request.

\bibliographystyle{apsrev4-2}
\bibliography{khalil_clean}

\end{document}